\documentclass[jou]{apa}
\usepackage{graphicx}
\usepackage{bm}
\usepackage{amsmath}
\usepackage{bbm}
\usepackage{amsfonts}
\usepackage{braket}
\usepackage{breqn}
\usepackage{xcolor, soul}
\sethlcolor{yellow}

\newcommand{\memvec}{\ensuremath{\mathbf{m}}}
\newcommand{\mem}{\ensuremath{{m}}}
\newcommand{\f}{\ensuremath{\tilde{f}}}

\newcommand{\covwhat}{\ensuremath{\Sigma_{\mathtt{what}}}}
\newcommand{\covwhen}{\ensuremath{\Sigma_{\mathtt{when}}}}

\newcommand{\E}{\ensuremath{\mathtt{E}}}

\newcommand{\rec}{\mathbf{R}}
\newcommand{\Recur}{\ensuremath{\mathbf{M}}}
\newcommand{\Mwhat}{\ensuremath{\Recur_{\textnormal{what}}}}
\newcommand{\Mwhen}{\ensuremath{\Recur_{\textnormal{when}}}}

\newcommand{\inh}{\ensuremath{\kappa}}
\newcommand{\ringact}{\ensuremath{r}}
\newcommand{\edgeact}{\ensuremath{r_E}}
\newcommand{\bumpact}{\ensuremath{r_B}}

\newcommand{\Wee}{\ensuremath{W_{EE}}}
\newcommand{\Wbb}{\ensuremath{W_{BB}}}

\newcommand{\spineedge}{\edgeact}

\begin{document}

\title{``What'' $\times$ ``When'' working memory representations using Laplace Neural Manifolds}

\fourauthors{Aakash Sarkar}{Chenyu Wang}{Shangfu Zuo}{Marc W.~Howard}

\shorttitle{What $\times$ when working memory}
\leftheader{Sarkar, Wang, Zuo and Howard}

\fouraffiliations{%
  Department of Psychological and Brain Sciences\\
  Boston University
}{%
  Department of Psychological and Brain Sciences\\
  Boston University
}{
	Department of Psychology\\
	Rutgers University
}{%
 Department of Psychological and Brain Sciences\\
 Department of Physics\\
 Boston University
}

\abstract{
Working memory---the ability to remember recent events as they recede continuously into the past---requires the ability to represent any stimulus at any time delay. This property requires neurons coding working memory to show mixed selectivity, with conjunctive receptive fields (RFs) for stimuli and time, forming a representation of ‘what’ x ‘when’ for the recent past. We study the properties of such a working memory in experiments where a single stimulus must be remembered for a short time. Conjunctive receptive fields allows the covariance matrix of the network to decouple neatly, allowing an understanding of the low-dimensional dynamics of the population. We study a specific choice---a Laplace space with exponential basis functions for time coupled to an ``Inverse Laplace'' space with circumscribed basis functions in time. We refer to this choice with basis functions that evenly tile log time as a Laplace Neural Manifold for time. Despite being related by a linear projection, the Laplace population shows a stable stimulus-specific subspace whereas the Inverse Laplace population shows rotational dynamics.  The rank of the covariance matrix with time grows logarithmically in good agreement with data. We sketch a continuous attractor network that constructs a Laplace Neural Manifold for time. The attractor in the Laplace space appears as an edge; the attractor for the inverse space as a bump. This work provides a bridge between abstract cognitive models of WM and circuit-level continuous attractor neural networks.
}


\maketitle

\begin{figure*}
\begin{center}
\includegraphics[width=0.96\textwidth]{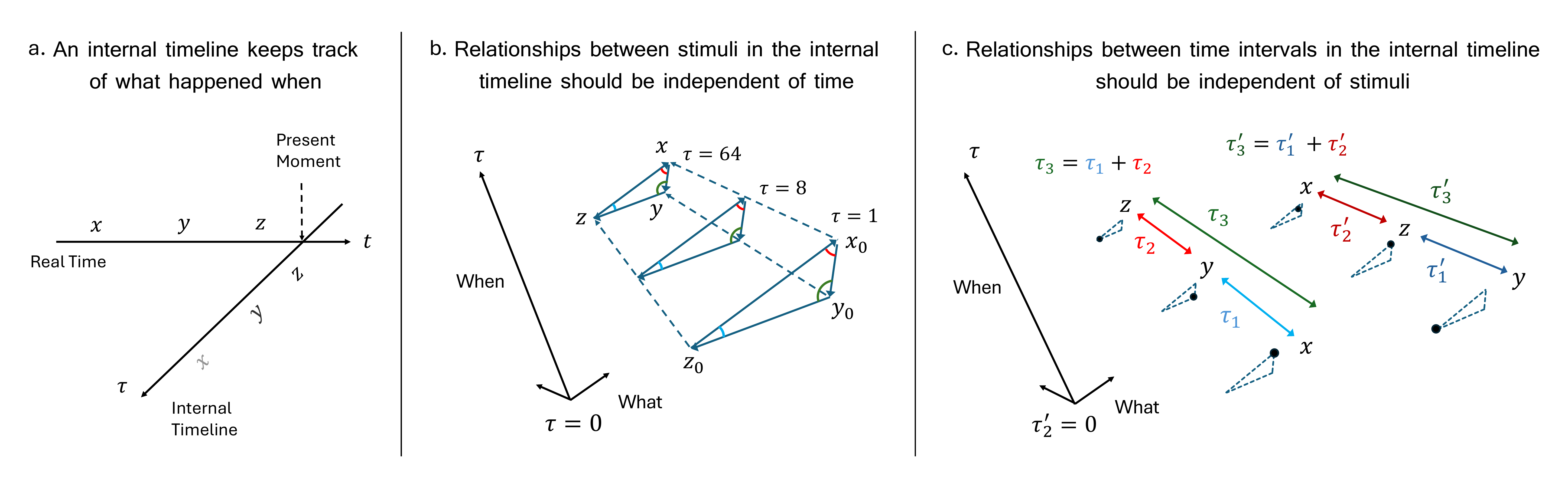}
\end{center}
\caption{
\emph{An internal timeline where stimulus (`What') representations change independently of temporal representations (`When'), and vice versa.}
\textit{Left: An internal timeline creating a record of the past.} We hypothesize that the brain forms a conjunctive representation that places events — a “what” — on an internal
timeline — a “when”. Such a conjunctive representation of what happened when should have properties, shown in the middle and right. \textit{Middle: The `What' representations should behave independently from the elapsed time, or `When'}. Consider we saw a set of stimuli, $x$, $y$, and $z$, at a delay of either $\tau =1$, $8$ or $64$ seconds. The representation vectors for stimuli $x$, $y$ and $z$ in our internal timeline might shrink or expand as we move into the past ($\tau$ gets bigger), but the relationships between them, here represented as the angle between each pair of vectors, should remain the same for all three durations ($\tau$), so that their meaning (`What') remains independent of the elapsed time (`When'). \textit{Right: The `When' representations should behave independently of the stimuli which are experienced.} We can represent different durations no matter what stimuli are used to define those intervals. For example, if three stimuli are presented in succession ($x$, $y$, $z$) such that the second one follows the first one by a delay of $\tau_1$ and the third one follows the second one by a delay of $\tau_2$, we are aware that the time interval between the first and third stimuli ($\tau_3$) is just the sum of the first two delays ($\tau_3=\tau_1+\tau_2$). This would hold true even if the stimuli (`What') were presented in a different order ($y$, $z$, $x$) - we might denote the time intervals differently ($\tau'_1$, $\tau'_2$, and $\tau'_3$), but the relationship between them (`When') would be unchanged ($\tau'_3=\tau'_1+\tau'_2$).
}
\label{fig:weylschem}
\end{figure*}

\begin{quotation}
\noindent The fact that Time is a form of our stream of experience is
expressed in the idea of equality: the empirical content which fills the
length of Time AB can in itself be put into any other time without being in
any way different from what it is.
	\flushright{--- Hermann Weyl, \emph{Space, Time, Matter}}
\end{quotation}
\nocite{Weyl22}

Consider a simple experiment where a single stimulus is presented for a brief moment followed by an unfilled delay interval.  With the passage of time, the memory for the event is preserved.  According to philosophers \cite{Jame90,Huss66,Berg10}, the identity of the stimulus in memory is unchanged, but the memory takes on a new character with the passage of time.  In the words of \citeA{Huss66}, with the passage of time, ``points of temporal duration recede, as points of a stationary object in space recede when I {`go away from the object'}.''
Experimental data from cognitive psychology is consistent with this introspection; participants can \textit{separately} judge the occurrence and relative time of different stimuli \cite{Hack80,Hint10}.

The great mathematical physicist Hermann Weyl introduced the argument above
(with analogous requirements for space) as part of an axiomatic derivation of
general relativity.  Viewed in the light of contemporary discussions about
artificial intelligence, we might say that Weyl required a compositional
representation of empirical content and time.  Given a way to describe
empirical content---a what---it must be possible to describe every possible
what at every possible when, \textit{and} vice versa. 

The way the early visual system conjunctively codes for what and
\emph{where} information provides a template for how the brain could code for
what and \emph{when} information.  
Throughout the early visual system there are neurons with similar sensitivity
to patterns of light, but with receptive fields in different locations over
retinal coordinates.  For instance, simple cells respond best to patterns of
light with a particular orientation---a what---at a particular location---a
where.  The pattern of activity of simple cells over all spatial locations
gives a conjunctive representation of what is where in the visual field.

Like position in the visual field, physical time is also an ordered continuous
dimension.  In much the same we might notice that object \textsc{a} is located
to the left of object \textsc{b}, we can also remember if event \textsc{a}
happened before or after event \textsc{b}.   This paper pursues the
implications of assuming a conjunctive representation that places events—--a
what--—on an internal timeline-—-a when---as a fundamental property of working
memory for the recent past.  

%
%
%


We show that neuronal populations with receptive fields of neurons as a a
product of a stimulus term and a temporal term satisfies the requirement of
compositionality of what and when \cite{MachEtal10}. Such conjunctive codes of
what $\times$ when make it straightforward to write out covariance matrices in
simple experiments, enabling a description of population dynamics in low
dimensional spaces like those widely used in neuroscience research. We then
specify a particular choice for temporal basis functions inspired by work in
theoretical neuroscience \cite{ShanHowa13}, cognitive psychology
\cite{HowaEtal15} and deep networks \cite{JacqEtal22}.  When projected onto a
linear space, these \emph{Laplace Neural Manifolds} for time generate
predictions that resemble empirical results from monkey cortex
\cite{MurrEtal17,CuevEtal20}. Finally, to illustrate that these equations can
be instantiated in biological networks, we sketch a continuous attractor
neural network model that conjunctively codes for what $\times$ when, with
temporal basis functions chosen as a Laplace Neural Manifold.

\section{Theoretical Considerations}

This paper looks into compositional representations and their implications and implementations in neuroscience. We show that a compositional working memory, coding for what happened when, requires that neurons must have conjunctive receptive fields, decomposable as functions of stimulus and time. Placing minimal restraints on the form of the receptive fields, we show that we can further write the population covariance as a tensor product of the stimulus ( $\covwhat$) and time ($\covwhen$) covariances, respectively. 

We simulate these populations for simple working memory tasks, where a task variable, distributed either on a ring or on a line, must be remembered for a short amount of time. Depending on our choice of temporal basis set, we observe, through linear dimensionality reduction techniques like PCA, neurally observed traits like stable subspaces and rotational dynamics.

Finally, since the covariance matrix of the population decomposes into  $\covwhat$ and $\covwhen$ - and the dimensionality of $\covwhat$ is completely fixed by the task and choice of stimulus RFs, we observe that the subspace spanned by $\covwhen$ controls the dimensionality of the covariance matrix. Measuring the dimensionality of neural trajectories as a function of time should reveal the density of basis functions over the continuous dimension of time. As a natural consequence of smooth basis functions, we see that the dimensionality of $\covwhen$ can grow without bound as a function of the elapsed time, but the growth can decelerate.

\begin{figure*}[tb]
\begin{center}
\includegraphics[width=0.95\textwidth]{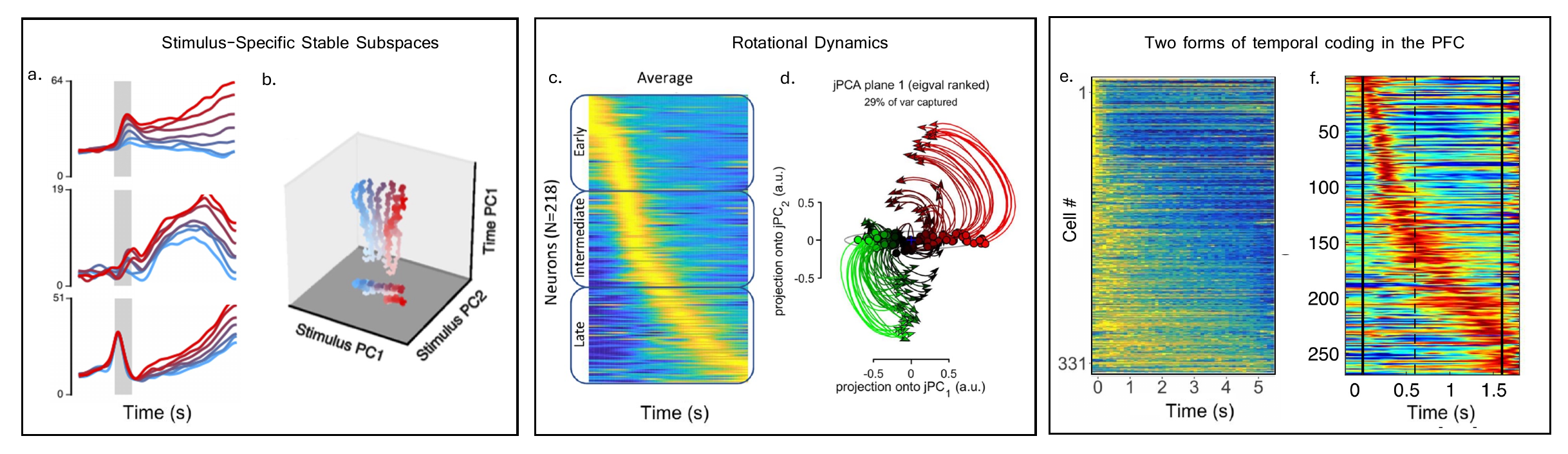}    
\end{center}
\caption{
\emph{Low-dimensional dynamics in neural populations maintaining a stimulus representation show specific trends like stable subspaces and rotational dynamics, while timing information is maintained using two different temporal coding schemes.}
\textit{Left: Stable subspaces exist simultaneously with diverse neural dynamics in Prefrontal Cortex (PFC) During Working Memory.} While classical models posited stable, persistent firing for working memory, individual PFC neurons exhibit highly heterogeneous and dynamic activity during delay periods \textit{(a)}. However, population-level analysis can reveal stable representations. For instance, in a vibrotactile delayed discrimination task, where one has to keep track of a frequency, Principal Component Analysis (PCA) of PFC neuronal populations uncovers a stimulus-specific subspace \textit{(b)}. Neural trajectories within this subspace maintain a stable representation of the task variable in the shape of a line, with temporal variance relegated to an orthogonal subspace (perpendicular to the stimulus subspace), with different colored lines corresponding to different values of the stimulus frequency. Adapted from \protect\citeA{MurrEtal17}. \textit{Middle: Rotational Dynamics in Motor Cortex.} In contrast to the stable representations in working memory, studies on motor control have revealed different dynamics. Churchland et al. (2012) observed rotational dynamics in neuronal populations of the motor cortex during reaching tasks, where trajectories in an informative subspace show an angular rotation over time \textit{(d)}. Such oscillatory dynamics have been proposed as a fundamental form of neural processing. However, more recent work suggests these rotational dynamics could also be a signature of consistent sequential neuronal activity in the analyzed data \textit{(c)}. Adapted from \protect\citeA{LebeEtal19}. \textit{Right: Two complementary forms of temporal coding in the prefrontal cortex (PFC).}  Temporal context cells (\textit{e}, adapted from \protect\cite{CaoEtal24}) exhibit persistent activity throughout a delay period, with their firing rate modulated by the passage of time. This suggests they maintain a stable representation of the current temporal context. Time cells (\textit{f}, adapted from \protect\cite{TigaEtal18a}) on the other hand, show transient bursts of activity at specific, successive moments within a trial, effectively tiling the delay period with their firing. }
\label{fig:data}
\end{figure*}

\subsection{Conjunctive receptive fields can support a compositional working memory}

Let us assume that we prepare an experiment in which one of several stimuli
$x$, $y$, $z$ is presented at $t=0$ for many trials.  We counterbalance the
number of presentations of each stimulus and perform all other experimental
controls.  
We record the firing rate over a population of neurons $\memvec$ that
reflects the state of a memory with finite duration  \cite{MaasEtal02}.
We choose the time between trials to be long enough that we can ignore any
carryover from previous trials.  Let us describe the population vector
expected at a time $t$ after presentation of a particular stimulus $\hat{x}$ as
$\memvec\left(\hat{x},t\right)$.

We operationalize Weyl's requirement that the empirical content of the
memory---the what---be unchanged with the passage of time by requiring that
$x$ be linearly decodable without knowing $t$.   It is acceptable that the
\emph{accuracy} of decoding changes with the passage of time.  However, in
order for the ``empirical content'' to remain fixed we require that the
relationships between all pairs of stimuli in the decoding space are
preserved at all time points (Fig.~\ref{fig:weylschem}b).

Just like the relationships in the what representations should
stay independent of the duration when they were experienced, the when
representations should also be independent of the identity of the
stimulus. The relationships between pairs of time intervals should thus
be preserved independent of the stimuli used to mark those intervals
(Fig.~\ref{fig:weylschem}c). If three stimuli are presented in succession
($x$, $y$, $z$) such that the second one follows the first one by a delay of
$\tau_1$ and the third one follows the second one by a delay of $\tau_2$, the
time interval between the first and third stimuli ($\tau_3$) is just the sum
of the first two delays ($\tau_3=\tau_1+\tau_2$). This relationship between
the intervals would not change even if the order of the stimuli themselves are
shuffled ($y$, $z$, $x$).

This requirement leads to the conclusion that neurons in
$\memvec\left(\hat{x},t\right)$ have
conjunctive receptive fields as a function of what $\times$ when.  
To see this, let us refer to the column of a linear decoder pointing in the direction
of $x$ as $\mathbf{x}$  and
define $\mem_x(\hat{x},t) = \mathbf{x}'\memvec\left(\hat{x},t\right)$ and
$\mem_y(\hat{x},t) = \mathbf{y}'\memvec\left(\hat{x},t\right)$.  
It is acceptable that the magnitude of $\mem_x\left(\hat{x},t\right)$ and 
$\mem_y\left(\hat{x},t\right)$ change as a function of time as information
recedes into the past.  However for the ``empirical content'' to be the same, we require that the relationships between all stimuli be constant.  This requires that $\frac{\mem_x\left(\hat{x},t\right)}{\mem_y\left(\hat{x}, t\right)}$ be constant as a function of time for all pairs of stimuli $x$, $y$ and all stimuli that could be presented $\hat{x}$.  For each cell in $\memvec$ projecting into $\mathbf{x}$ with a particular time course, there must be another cell projecting into $\mathbf{y}$ with the same time course, up to multiplication by a constant.   

We can thus subscript the cells in the population vector
$\memvec$ with two indices $i$ and $j$
\begin{eqnarray}
	\mem_{ij}\left(\hat{x},t\right) &=& a_{ij}\ {g}_i\left(\hat{x}\right) {h}_j(t)
	\label{eq:memgh}
\end{eqnarray}
where $a_{ij}$ is just a normalization factor that we will set to $1$ for simplicity.
Note that this decomposition of $\memvec\left(\hat{x},t\right)$ also satisfies the other constraint to the effect that we can also linearly decode the when, $t$,  without knowing the what, $\hat{x}$.

Conjunctive, mixed selective receptive fields can create a compositional representation of what and when. We can understand $g_i(\hat{x})$ as a set of receptive fields over whatever dimensions span the stimulus space and $h_j(t)$ as a set of temporal receptive fields.  

This requirement would not be compatible with many forms of memory, some of which are widely used in neuroscience.  For instance, suppose that $\memvec\left(\hat{x},t\right)$ were maintained as a function of time with an RNN initialized as $\memvec\left(\hat{x}, t= 0\right) = \mathbf{I}(\hat{x})$ and that then evolved as $\dot{\memvec} = \rec \memvec$. Being able to decompose $\memvec\left(\hat{x},t\right)$ as Eq.~\ref{eq:memgh}
requires that $\mathbf{R}$ be decomposable as 
\begin{equation}
\rec = \Mwhat \otimes \Mwhen
\label{eq:RNNconst}    
\end{equation}
and that $\mathbf{I}(x)$ provides the same input to each part of the space
spanned by $\Mwhen$ for each stimulus. This is a very strong
constraint on recurrent connectivity that reflects a specific structural
inductive bias.  This choice is not typically made in computational
neuroscience network models. 






\subsection{Total covariance $\Sigma$ decouples into \covwhat \hspace{0.2pt}  and \hspace{0.2pt} \covwhen}

To understand the behavior of neural data, linear dimensional reduction
techniques is often used to visualize population vectors as neural trajectories
in time. This requires knowledge of the population covariance matrix of the
neural data.  If the activity of populations of neurons can be decomposed into products of 
what $\times$  when receptive fields, as in Eq.~\ref{eq:memgh}, then this results in 
straightforward expressions for the population covariance matrices.

Lets start with a compositional representation of what happened when, with neurons obeying
\begin{equation}
	\Phi_{ij}\left(\hat{x},\tau\right) = g\left(\hat{x},x_i\right)\,h(\tau,\tau_j) \label{eq:1}
\end{equation}
where the activity $\Phi$ of the neuron (indexed by a \textit{what} index  $i$ and a \textit{when}
index $j$) describes the receptive field when stimulus $\hat{x}$ is a time $\tau$ in
the past.  If a stimulus is presented at time $t=0$, then at time $t$, the
stimulus is $\tau = t$ seconds in the past.  The
function $g$ describes the neuron's tuning curve over the stimulus dimension,
with the preferred stimulus $x_i$, and the function $h$ describes the temporal
receptive field of the neuron, with a time sensitivity parameter $\tau_j$. The
stimulus and time preferences can be indexed separately, since the choice of one
has no bearing on the other.

Let us calculate the general covariance matrix for populations that obey \eqref{eq:1} for some generic set of stimuli indexed by $x$. We are not restrained to any specific choice of receptive fields for $g$ and $h$, but only make the assumption that they are normalized over the stimulus and time space respectively
\begin{eqnarray}
    \frac{1}{T} \int_0^T h(\tau_j,\tau)\, d\tau = 1,  \quad \quad 
    \frac{1}{N} \,  \sum_{x} g(\hat{x},x_i) = 1.
	\label{eq:norm}
\end{eqnarray}
This allows us to write the expectation of the activity over time as a pure
function of stimulus, $\langle\Phi_{ij}\rangle_t(\hat{x}) \equiv \E_t[\Phi_{ij}](\hat{x})
=  g(\hat{x},x_i)$, and the expectation of the activity over stimulus as a
pure function of experimental time $\langle\Phi_{ij}\rangle_{\hat{x}}(t) \equiv\E_{\hat{x}}[ \Phi_{ij}](t)
= h(\tau_j,\tau=t)$.

For \covwhen, the covariance over time of $\langle\Phi_{ij}\rangle_x(t)$ and $\langle\Phi_{kl}\rangle_x(t)$,
we get 
\begin{equation}
 [\covwhen]_{\,ijkl}
= \frac{1}{T}\int_0^T h(\tau,\tau_j) \, h(\tau,\tau_l)\,d\tau - 1. 
\end{equation}
Note that the first term can be written as $1_{ik}\,H_{jl}$, where
$\mathbf{H}$ is a symmetric matrix which encodes the expectation (over time)
of the product of $h(\tau,\tau_j)$ and $h(\tau,\tau_l)$.

For $\covwhat$, the covariance over stimulus of $\langle\Phi_{ij}\rangle_t(\hat{x})$ and $\langle\Phi_{kl}\rangle_t(\hat{x})$, we get
\begin{equation}
[\covwhat]_{\,ijkl}
= \frac{1}{N} \sum_x g(\hat{x},x_i)\,g(\hat{x},x_k) - 1.
\end{equation}
Note that the first term can be written as $G_{ik}\,1_{jl}$ where $\mathbf{G}$ is a symmetric matrix which encodes the expectation (over stimuli) of the product of $g(\hat{x},x_i)$ and $g(\hat{x},x_k)$.

The overall covariance  $\Sigma_{\,ijkl}$  is calculated over both stimuli and time and can be written as

\begin{multline*}
\Sigma_{\,ijkl} + 1 = \left( \frac{1}{N} \sum_x g(\hat{x},x_i)\,g(\hat{x},x_k) \right)\\
\times\left( \frac{1}{T}\int_0^T h(\tau,\tau_j)\, h(\tau,\tau_l)\,d\tau \right)
\end{multline*}
where the second term on the left appears as the expectation of the activity over both stimulus and time and is unity since $g$ and $h$ are normalized
appropriately. For the first term, the sum
and integral can be separated since $g$ and $h$ are pure functions of stimulus
and time. Recognizing that the first term is just a product of the matrix
elements $G_{ik}$ and $H_{jl}$, we can generalize this to 
\begin{equation}
\mathbf{\Sigma} =  \mathbf{G} \otimes \mathbf{H} - \mathbf{1}.
\end{equation}
We can thus write the overall covariance matrix ($\Sigma$) as a Kronecker product 
\begin{equation}
	\mathbf{\Sigma} + \mathbf{1} =  \left(\mathbf{\Sigma}_{\text{what}} + \mathbf{1}\right) \otimes \left(\mathbf{\Sigma}_{\text{when}} + \mathbf{1}\right).
    \label{eq:covproduct}
\end{equation}
Thus, we see that the overall covariance matrix $\Sigma$ (computed over both
time and stimulus dimensions) neatly decomposes into the tensor product of the
covariance matrices for stimulus ($\Sigma_{what}$) and time ($\Sigma_{when}$)
respectively, due to the conjunctive coding designed into the Laplace neural
manifolds. 



\begin{figure*}[tb]
\begin{center}
\includegraphics[width=0.95\textwidth]{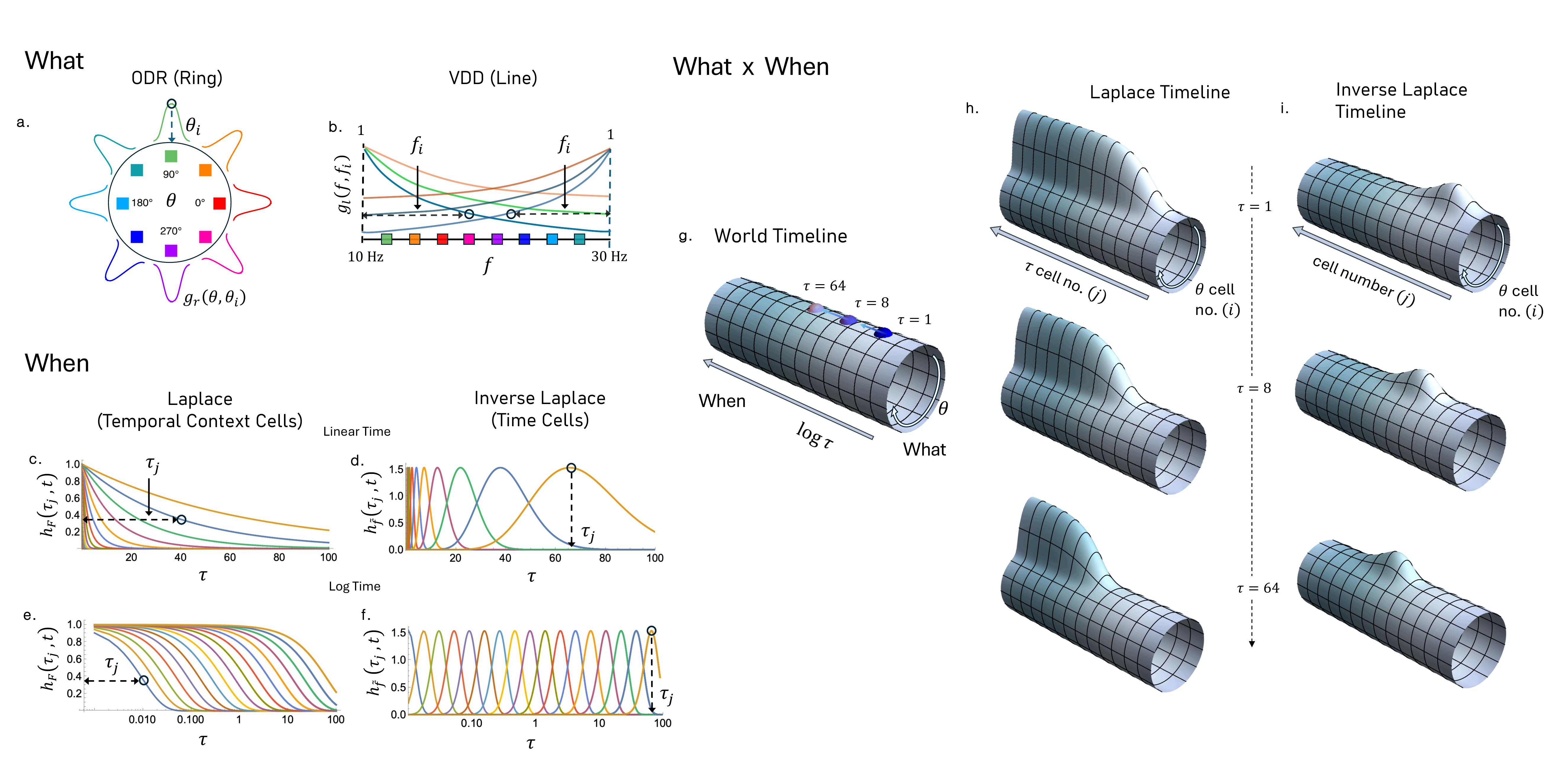}    
\end{center}
\caption{
\emph{Conjunctive coding using Laplace Neural Manifolds.}  LNM neurons have receptive fields (RFs) which are a product of a stimulus (\textit{what}) and a temporal (\textit{when}) term. 
\textit{What:} The stimulus receptive fields tile the stimulus space, which is a ring in the case of ODR (Oculomotor Delayed Response, where an angular location $\theta$ has to be remembered), and a line in the case of VDD (Vibrotactile Delayed Discrimination task, where a frequency $f$ has to be remembered). For the ODR task, neurons have bell-shaped tuning curves approximating a circular analogue of a normal distribution, preferentially encoding stimuli at different angles $\theta_i$ which evenly tile the ring(\textit{a}). For the VDD task, the frequency range from $10$ to $30$ $Hz$ is tiled by exponentially decaying and ramping cells with a spectrum of different ramp rates $1/f_i$. (\textit{b}). \textit{When:} Laplace cells ($F$) decay exponentially with different decay rates $1/\tau_j$, after the stimuli is presented, resembling temporal context cells (\textit{c}), while inverse Laplace ($\tilde{f}$) cells fire sequentially, peaking at different times $\tau_j$, resembling time cells (\textit{d}). The receptive fields are chosen to evenly tile log time \textit{(d, e)}. When a stimuli is presented at a certain angle $\hat{\theta}$, the activity of the population as the time $T$ after the presentation can be modeled on a cylinder \textit{(e)}. Laplace neurons (modeling temporal context cells) encode this history as a moving edge when seen in log time \textit{(f)}, while the inverse Laplace neurons encode it as a bump \textit{(g)}.}
\label{fig:LMNs}
\end{figure*}

\begin{figure*}[tb]
\begin{center}
\includegraphics[width=0.8\linewidth]{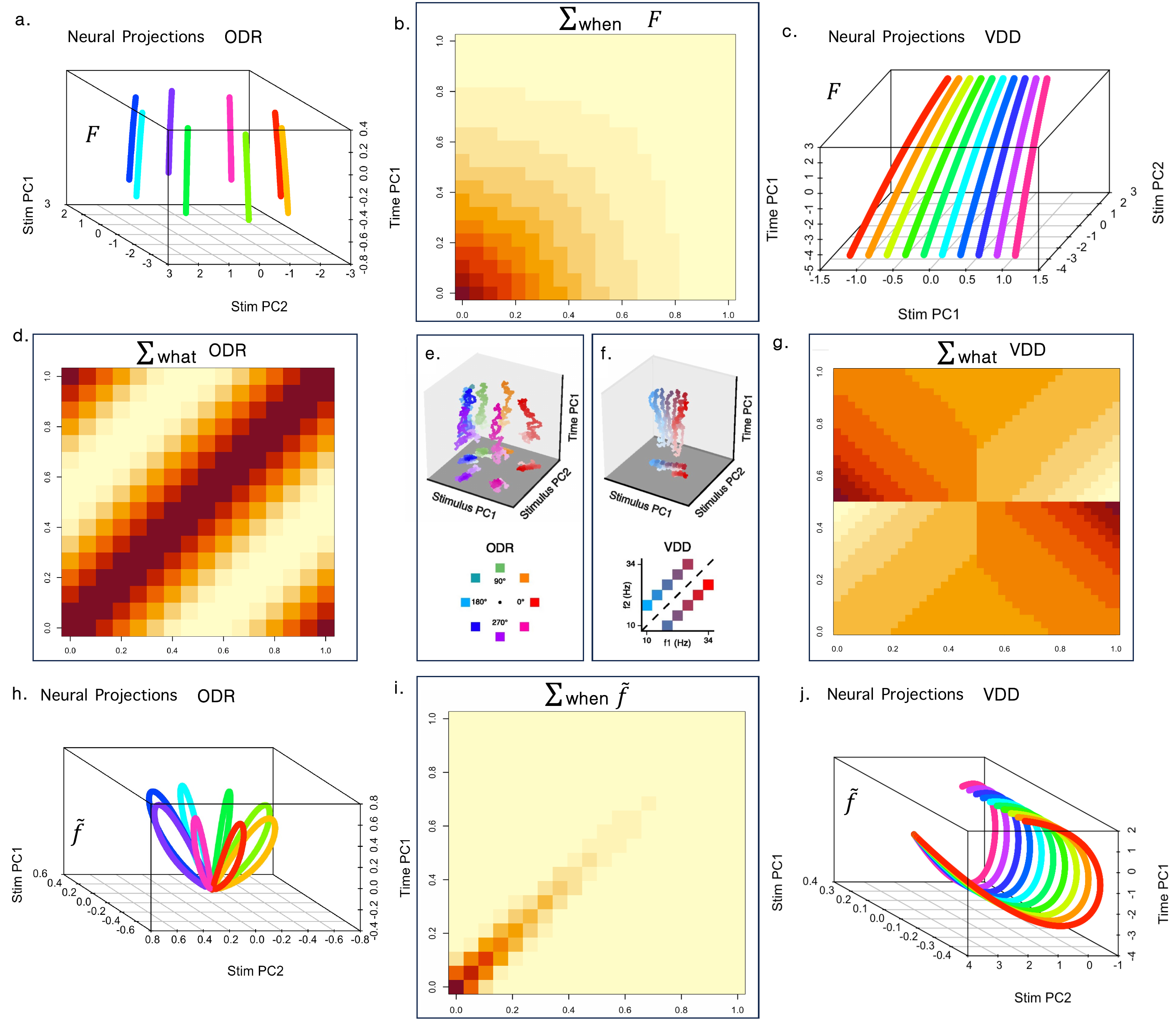}
\end{center}
\caption{
\emph{Neural trajectories from LNMs can show stable coding as well as temporal dynamics.}
The figures show neural trajectories for ODR  (Left) and VDD (right) tasks, computed with Laplace Neural Manifold populations simulating temporal context cells (Laplace, $F$, Top row, \textit{a} and \textit{c}) and time cells (inverse Laplace, $\tilde{f}$, Bottom row, \textit{h} and \textit{j}) respectively. The neural activity is projected, via eigendecomposition of the covariance matrices,  onto the first two stimulus principal components (Stim PC1 and PC2, $x$ and $y$ axes) and the first temporal principal component (Time PC1, $z$ axes), up to a sign factor. Neural trajectories corresponding to different initial stimulus conditions, the angle $\theta$ for the ODR task, shown on the left (\textit{a} and \textit{h}), and the frequency $f$ for the the VDD task (Right, \textit{c} and \textit{j}), are represented with different colors. The overall covariance matrix of the populations can be understood as a tensor product of covariance matrices across different `whats' -  stimulus representations for the ODR (\textit{d}), and VDD tasks (\textit{g}), and different `whens' - temporal coding representations corresponding to the Laplace cells $F$ (\textit{b}), and the inverse Laplace cells $\tilde{f}$ (\textit{i}). The corresponding primate PFC neural trajectories, adapted from \protect\citeA{MurrEtal17}, are shown for the ODR task (\textit{e}) and VDD tasks (\textit{f}) \textit{(Top)}, along with the stimulus spaces for both tasks \textit{(Bottom)}.}
\label{fig:Murray}
\end{figure*}

\section{Low-dimensional projections of What $\times$ When conjunctive representations}

A comparison with experimental data requires us to specify the form of the receptive fields for what and when information. For concreteness, we study reasonable choices of temporal receptive fields and examine working memory population dynamics during simple delay experiments that have been used to study cortical populations in monkeys.  The choices of stimulus receptive fields are appropriate to the stimuli used in each of two
tasks.  The temporal receptive fields are chosen consistent with two broad
classes - each choice forms a basis set over the time dimension, and they are
provably related to one another by a linear operator. We will see
that these two closely related choices of temporal basis functions lead to
qualitatively different population dynamics as a function of time when
projected onto a low-dimensional space.

\subsection{Laplace Neural Manifolds: Two forms of temporal basis functions}
We choose two forms of receptive fields,
both of which form a basis set over the continuous dimension of time.
One set of basis functions is chosen such that each cell fires in a circumscribed region of time.  As a ``what'' recedes into the past, cells
tiling the time axis fire sequentially, much like so-called time cells  (Fig.~\ref{fig:data}f) that
have been observed in the hippocampus and elsewhere \cite{PastEtal08,JinEtal09,MacDEtal11}. Although cells that sequentially activate in time are widely observed in the brain \cite{TigaEtal18a,AkhlEtal16,ParkEtal22,SubrSmit24}, these are not the only choice of basis functions.

In addition, we consider monotonic receptive fields in which individual
neurons decay as a function of time, but with different time
constants (Fig.~\ref{fig:data}e) \cite{TsaoEtal18,BrigEtal20,ZuoEtal23,CaoEtal24,AtanEtal23}. 
Note that when projected into principal component space, decaying firing is
indistinguishable from ramping firing. 

\paragraph{Exponential temporal receptive fields for Laplace transform.}
For exponential receptive fields, we set the function $h$ as
\begin{equation}
	h_{F} (\tau,\tau_j) \propto e^{-\frac{\tau}{\tau_j}},
	\label{eq:h_F}
\end{equation}
where the constant of proportionality is set to ensure the normalization of
$h$ specified in Eq.~\ref{eq:norm}.  Comparison
of these receptive fields with the definition of the Laplace transform shows
that at time $t$ this population is closely related to the Laplace transform
with real coefficients, of a delta function a time $\tau$ in the past. This
justifies the claim that $h_F$ is a basis set if $\tau_j$ is effectively
continuous. 
We refer to the population with temporal receptive fields $h(\tau_j, t) =
e^{-t/\tau_j}$ as a Laplace population and label this population as $F$  (Fig.~\ref{fig:LMNs}c).

\paragraph{The inverse space: Circumscribed temporal receptive fields.}
To construct circumscribed temporal receptive fields, we use gamma functions
\cite{TankHopf87,deVrPrin92}
parameterized by a variable $k$
\begin{equation}
h_{\f} (\tau_j,\tau) =  \left(\frac{\tau}{\tau_j}\right)^{k} e^{-k(\tau/\tau_j)}.
	\label{eq:gamma}
\end{equation}
This receptive field is a product of a power law  that grows with $\tau/\tau_j$
and an exponential that decays with $\tau/\tau_j$.
With this choice, $h(\tau,\tau_j)$ goes to zero as $\tau/\tau_j \rightarrow 0$ and
also as $\tau/\tau_j \rightarrow \infty$.  In between there is a single peak at a time
that depends on the choice of $\tau_j$.  Choosing a new value of $\tau_j$
rescales the function.  Thus a population with different values of $\tau_j$
will fire sequentially as a triggering stimulus enters the past
(Fig.~\ref{fig:LMNs}d).

It can be shown that the choice of temporal receptive fields in
Eq.~\ref{eq:gamma} leads to a deep relationship with neurons with
receptive fields described by Eq.~\ref{eq:h_F}.  Whereas a population
of neurons with exponential receptive fields and a variety of time constants
encode the Laplace transform with real coefficients, a population of neurons
with receptive fields obeying Eq.~\ref{eq:gamma} approximate the inverse
Laplace transform, computed using the Post approximation with coefficient $k$
\cite{ShanHowa12}.  This means that the two populations are related to one
another \emph{via} a linear transformation that can readily be computed
\cite{ShanHowa13}.  For this reason, we refer to a population with receptive
fields chosen as Eq.~\ref{eq:gamma} as an Inverse Laplace Space and label
the activity across the population as $\f$ .

\subsection{Distributions of time constants for log time}
Each Laplace neuron has a time constant $\tau_j$ that dictates how fast its
activity decays, while each Inverse Laplace neuron has a time constant
$\tau_j$ which dictates the position of its temporal receptive field. 
It remains to choose the distribution of time constants $\tau_j$ for the
populations.  Rather than a basis set uniformly tiling $\tau$, we select time
constants so that the basis functions tile $\log \tau$.  This choice has
theoretical advantages  \cite{WeiStoc12,Pian16,HowaShan18} and is in agreement
with some neural data \cite{CaoEtal22,GuoEtal21}.

The distribution of the time constants $\tau_j$ is instrumental in
defining how the basis functions sample time.  In this paper, the $\tau_j$ are distributed
geometrically.
The time constants for the temporal receptive fields 
are chosen such that the $n$th time constant is given by \begin{equation}
    \tau_n = \left(1 + c\right)^{n}\ \tau_0
	\label{eq:log}
\end{equation} 
for some positive constant $c$. If we rearrange the terms a bit, we can express $n(\tau)$, the number of cells that tile the time axis from 0 to $\tau$
\begin{equation}
    n(\tau) = log_{1 + c}\ \frac{\tau}{\tau_0}.
	\label{eq:nlog}
\end{equation} 
The number of cells spanning an interval $\tau$ thus grows logarithmically with $\tau$. This is also equivalent to saying that
the time constants evenly tile  $\log$ time
\begin{eqnarray}
    \log{\tau_{n+1}} - \log{\tau_{n}}= \left(1 + c\right).
\end{eqnarray}

(Fig.~\ref{fig:LMNs}b,c) shows the activity of cells with two kinds of temporal
receptive fields firing as a function of experimental time, and as a
function of log time (Fig.~\ref{fig:LMNs}d,e).  We will refer to these paired sets of basis functions with
time constants chosen as in Eq.~\ref{eq:log} as a \textit{Laplace Neural Manifold}
\cite{DaniHowa25,HowaEtal24}.


\subsection{`What' Representations}
The experimental data reported in \citeA{MurrEtal17} used simple
memory tasks using two different types of to-be-remembered stimuli.
The oculomotor delayed response (ODR) task requires a monkey to remember the
angle of a visual stimulus for a later saccade.
The vibrotactile delayed discrimination (VDD) task requires a monkey to remember
the frequency of a vibrotactile stimulus for a short time and then compare it
to the frequency of a test stimulus.  We model the stimulus receptive fields
$g(x_i, \hat{x})$ in these two experiments as a ring describing the possible
angles and a line describing log frequency of the vibrations respectively
(Fig.~\ref{fig:LMNs}a,b).  

For the ODR task (\textit{Ring}), the stimulus specificity of each cell is specified by a symmetric tuning curve function $g(\theta_i,\hat{\theta})$, where $\theta_i$ is preferred angle of cell $i$ and $\hat{\theta}$ is the stimulus presented in the experiment. We choose the tuning curves $g$ to follow the von Mises distribution, which serves as a close approximation to the wrapped normal distribution over a circle, normalized approximately. Cells are chosen so that their preferred directions $\theta_i$ tile the stimulus space uniformly, and simulated for eight stimuli conditions $\hat{\theta}$ evenly distributed on the ring (Fig.~\ref{fig:LMNs}a). 

For the VDD task (\textit{Line}), the stimulus dimension is along a line, and ten stimulus conditions $\hat{f}$ are chosen which tile this frequency space evenly from a minimum frequency of $f_{\tt{min}}=10\, \textnormal{Hz}$ to a maximum frequency $f_{\tt{max}}=30\,\textnormal{Hz}$. The tuning curves coding this space are chosen to be decaying and ramping cells, clamped at these minimum and maximum frequencies, respectively, with rate constants $1/f_i$ distributed geometrically, such that $f_i$ tile the frequencies between $f_{\tt{min}}$ and $f_{\tt{max}}$ logarithmically (Fig.~\ref{fig:LMNs}b), akin to the choice of time constants for `When' representations.

\paragraph{Visualizing conjunctive representations.}
Fig.~\ref{fig:LMNs}g-i provides a way to visualize the activity over the
network for stimuli chosen on a ring at different time points.  Each stimulus
presented at the beginning of a trial can be described by an angle.  As the
stimulus recedes into the past, the ``true past'' that would be recorded by a
perfect observer with, say, a video camera is describable as a point on a
cylinder.  Fig.~\ref{fig:LMNs}g provides a cartoon of this ``true past'' at
three moments in external time as a function of $\log \tau$.   In this
depiction, the present is closest to the viewer.

Fig.~\ref{fig:LMNs}h shows the pattern of activity over the conjunctive what
$\times$ when representation with Laplace temporal receptive fields as a
function of temporal index $n$.  Fig.~\ref{fig:LMNs}i shows the conjunctive
representation with Inverse Laplace temporal receptive fields displayed in the
same way.  For the Laplace representation, the angle of the stimulus controls
the angular location of the edge.  As time passes and the stimulus
recedes into the remembered past, the edge moves as a function of $\log \tau$.
The Inverse Laplace space has the same properties, except that the remembered time
is represented as a bump of activity. Memory for different kinds of stimuli would require a different structure along the stimulus directions.

\subsection{Dynamics of conjunctive representations in low-dimensional linear
projections}

We study the population dynamics of conjunctive representations of what
$\times$ when with Laplace and Inverse Laplace temporal receptive fields in
both ODR and VDD tasks. Analyzing state-space population trajectories of
primate PFC neurons, \citeA{MurrEtal17} found stimulus-specific subspaces with
persistent `What' representations that held stable even as individual neurons
exhibited heterogeneous temporal dynamics. 
We employ a similar methodology to look at the projections of
population-level activity, using Principal Component Analysis (PCA) to
account for variance, separately over the stimulus and time space, to generate
neural trajectories $\boldsymbol{\varphi}(\hat{x},t)$.

\paragraph{Projecting neural trajectories onto stimulus and time Principal Components.}
To calculate PCA over stimuli, we compute the stimulus covariance $\covwhat$
using the time-averaged delay activity $\langle\Phi_{ij}\rangle_t(\hat{x})$,
and use eigendecomposition to extract the first two principal axes
$\mathbf{v}_{\hat{x},1}$ and $\mathbf{v}_{\hat{x},2}$. The projection of the
mean-subtracted population activity onto the two principal axes gives us the
stimulus principal components 
\begin{equation}
\varphi_k(\hat{x},t) = \mathbf{v}_{\hat{x},k}
	\cdot\left[\Phi_{ij}\left(\hat{x},t\right) -
	\langle\Phi_{ij}\rangle_{\hat{x},t}\right],\quad \textnormal{for } k=1,2.
\end{equation}
This gives us the first two axes of our neural trajectories. The $z$ axis is
constructed to be orthogonal to the stimulus subspace defined by the stimulus
axes $\mathbf{v}_{\hat{x},k}$, which captures the largest component of
time variance in the neural activity. To define this, we calculate the time
covariance $\covwhen$ over the stimulus-averaged activity
$\langle\Phi_{ij}\rangle_{\hat{x}}(t)$, and extract the first principal axis
$\mathbf{v}_{t,1}$. We then orthogonalize this axis to the stimulus subspace
by subtracting the stimulus principal axes from it and normalizing

\begin{equation}
\mathbf{v}'_{t,1} = \frac{\mathbf{v}_{t,1} - \mathbf{v}_{\hat{x},1} -  \mathbf{v}_{\hat{x},2}}{\lVert \mathbf{v}_{t,1} - \mathbf{v}_{\hat{x},1} -  \mathbf{v}_{\hat{x},2}\rVert}.
\end{equation}
We now project the mean subtracted population activity onto the orthogonalized time axis to get the $z$ axis of the neural trajectories
\begin{equation}
\varphi_3\left(\hat{x},t\right) = \mathbf{v}'_{t,1}
	\cdot\left[\Phi_{ij}\left(\hat{x},t\right) -
	\langle\Phi_{ij}\rangle_{\hat{x},t}\right].
\end{equation}
There now exist neural trajectories $\boldsymbol{\varphi}$ corresponding to
each stimulus condition $\hat{x}$, which are constructed to have two axes which
captures stimulus variance, and a $z$ axis that captures time variance (Fig. \ref{fig:Murray}).  

\subsubsection{Laplace cells, resembling temporal context cells, show stimulus-specific stable subspaces} Laplace cells, which have exponentially decaying temporal receptive fields, generate a covariance matrix with smoothly decaying activation along the diagonal (from bottom left to top right), which is the direction of maximum variance  (Fig.~\ref{fig:Murray}b). The first principal component picks this up to show a monotonically varying trend as well (Suppl. Fig.~\ref{fig:supp}, c2). 

When the neural activity of Laplace cells is projected onto this smoothly and monotonically varying temporal principal component ($z$ axis), the stimulus representation thus remains stable, with only a relative scaling of the relative distances between the trajectories for different stimuli. Thus, for both ODR (Fig.~\ref{fig:Murray}a) and VDD (Fig.~\ref{fig:Murray}c) tasks. Although the populations show strong temporal dynamics and show a heterogeneity of timescales, taken together, the population-wide activity seems to encode a stable representation of the stimulus space.

\subsubsection{Inverse Laplace cells, resembling time cells, show rotational dynamics}
Inverse Laplace neurons, which have sequential receptive fields tiling a one-dimensional time-line, generate a mostly diagonal covariance matrix (Fig.~\ref{fig:Murray}i), decaying monotonically from bottom left to top right. The biggest principal components pick this diagonal direction to explain the most variance, and to tile this diagonal line end up being oscillating sinusoids (Suppl. Fig.~\ref{fig:supp}, d2) with a decaying envelope. When the neural activity is projected onto this oscillating component (\textit{z}-axis), the trajectories end up rotating in the low-dimensional space.  

The population trajectories for both ODR (Fig.~\ref{fig:Murray}h) and VDD (Fig.~\ref{fig:Murray}j) tasks thus show rotational dynamics, although the activity of the neurons themselves do not intrinsically have a oscillatory component. Sequential activity in the brain has also previously been  documented to produce artefactual rotational dynamics in linearly projected low-dimensional neural trajectories \cite{LebeEtal19,Shin23}.

Even though both populations share the same `what' representation, their activity, when projected onto the stimulus and time principal components, paints very different pictures of their low-dimensional dynamics. This happens due to the choice of temporal basis functions, even though the Laplace and Inverse Laplace representations are simply related by a linear transformation. Both temporal representations are high-rank - as evidenced by the variance in their covariance matrix \covwhen, explained by their principal components falling off smoothly (as seen in Suppl.\, Fig.~\ref{fig:supp}, c3 and d3). However, the nature of their receptive fields (ramping vs. sequences) gives rise to qualitatively different principal components for time and thus different linear low-dimensional dynamics like stable subspaces or oscillatory dynamics.

\begin{figure*}[tb]
\begin{center}
    \includegraphics[width=0.87\textwidth]{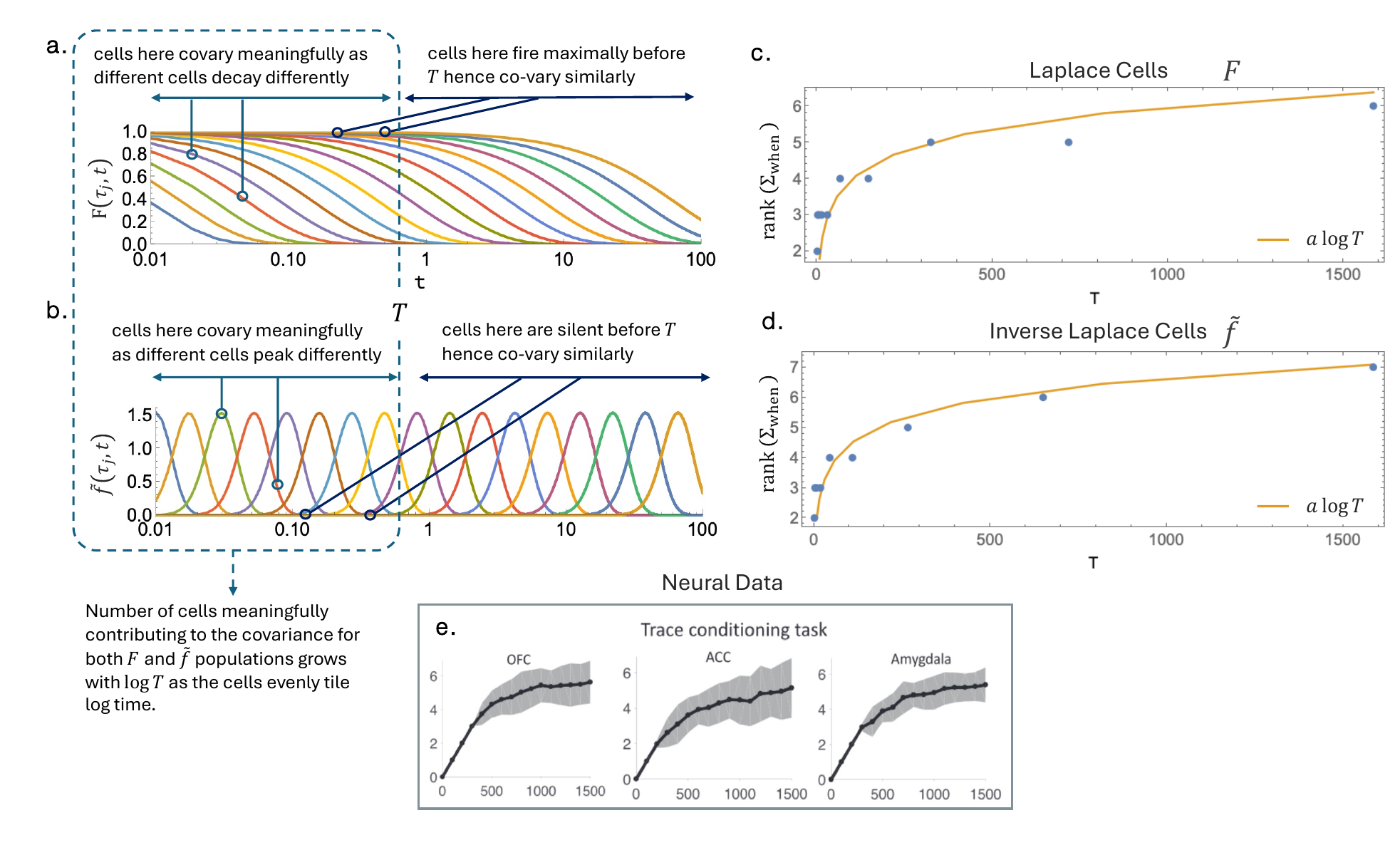}    %
\end{center}
\caption{
\emph{The cumulative dimensionality of neural trajectories in LNM cells grows
	logarithmically with elapsed time.} Both Laplace and Inverse Laplace
	cells tile log time evenly. When calculating covariance from $0$ to
	$T$, cells with time constants much larger than $T$ either both fire
	maximally (Laplace cells, \textit{a}) or are both silent (Inverse
	Laplace cells, \textit{b}). For both choices of temporal basis
	functions, only cells which have time constants less than $T$ co-vary
	meaningfully and contribute to the total covariance, and the number of
	such cells grows as $\log T$. Explicitly calculating the rank of the
	covariance matrix of simulated Laplace and Inverse Laplace cells, for
	different values of $T$, shows this empirically (\textit{c, d}), and
	gives us a measure of the dimensionality of their neural trajectories.
	This seems to follow the growth of cumulative dimensionality of actual
	neural data (\textit{e}) collected during WM tasks, adapted from \protect\citeA{CuevEtal20}.  \label{fig:Cueva}}
\end{figure*}

\section*{Dimensionality of neural trajectories grows with the distribution
of time constants}

In this section, we
examine the dimensionality of the neural space spanned by the population
trajectories of neurons with these temporal basis functions.  
For the choices of temporal basis functions chosen here
the dimensionality spanned by the network out to a particular time $T$ depends only on the
distribution of time constants. We compute the rank of the overall covariance matrix calculated using times from 0 to $T$  as a measure of dimensionality of neural trajectories.

\subsubsection{The rank of the total covariance grows with $\covwhen$}

We have seen in Eq.~\ref{eq:covproduct} that the total covariance can be
expressed as a tensor product of $\covwhen$ and $\covwhat$. 
Because $\covwhat$ is not a function of time, all of the time dependence of
the covariance matrix must be carried by $\covwhen$. This implies that the
growth of linear dimensionality of the space should be the same for different
kinds of stimuli maintained in working memory.
Starting from Eq.~\ref{eq:covproduct}, using inequalities describing ranks of matrix sums and products, it is possible to establish lower bounds on the rank of the total covariance

\begin{equation}
	\tt{rank}\left(\mathbf{\Sigma}\right) + 1  \geq
	\left[\tt{rank}\left(\mathbf{\Sigma}_{\tt{what}}\right) - 1 \right]
	\, \left[\tt{rank}\left(\mathbf{\Sigma}_{\tt{when}}\right) - 1\right].
\end{equation}
We can see that the rank of the the total covariance goes up with the rank of
$\covwhen$, as the rank of $\covwhat$ remains fixed once we choose a
particular experimental paradigm and the shape and number of receptive fields
tiling the stimulus space.   Thus, to determine the dimensionality of the
neural representation as a function of the total recording time $T$, we only need
to assess the rank of $\covwhen$ estimated from an experiment with the first
$T$ seconds of the delay.

Before stating the result, let us provide an intuition about how the rank of
$\covwhen$  should change as a function of $T$.
Consider a pair of Laplace cells (Fig.~\ref{fig:Cueva}a) with different time constants.  If
their time constants are less than $T$, they will have some covariance and
contribute to the rank of the covariance matrix. However, consider pairs of
cells that both have time constants much greater than $T$.  
These cells would both be firing maximally throughout the interval $0\leq t
\leq T$.  The covariance estimated up to time $T$ between these cells is
zero. Thus, the pair of cells with time constants much greater than $T$
would not affect the rank of covariance matrix. A similar argument can be made for the Inverse Laplace cells (Fig.~\ref{fig:Cueva}b).

\subsubsection{Geometrically spaced time-constants result in logarithmic growth in dimensionality of $\covwhen$ with recording time T}

As $T$ changes, the number of cells that contribute to the rank of the
covariance matrix goes up like the density of the basis $n(T)$, as defined in Eq.~\ref{eq:nlog}.  Thus, a logarithmic distribution of
time constants, as in Eq.~\ref{eq:log}, implies that the linear
dimensionality of the population measured from 0 to $T$  should grow like
$\log T$.  Fig.~\ref{fig:Cueva}c,d  show this argument corroborated
numerically as the rank of $\covwhen$, computed for both Laplace and Inverse Laplace neurons
grows logarithmically as a function of $T$.
This matches the empirical trend of growth of the cumulative dimensionality with time
(Fig.~\ref{fig:Cueva}e) recorded from cortical regions when performing WM tasks \cite{CuevEtal20}.

For generally high recording times, the dimensionality of neural trajectories thus approaches full rank for both Laplace and Inverse Laplace representations. These neurons show nonlinear mixed selectivity, which has previously been shown to be a signature of high-dimensional representations and enables a wide combination of linear readouts \cite{FusiEtal16}. Just like neurons in the monkey PFC show high shattering dimensionality \cite{PosaEtal24, BernEtal20} which allows a linear readout to disambiguate different stimulus experimental conditions effectively, Laplace and Inverse Laplace neuronal populations have a high-rank temporal covariance matrix, which allows a wide range of subsets of the continuum of time intervals to be linearly decoded.

\section*{Neural Circuit Model for Compositional Working Memory of What $\times$ When}

A compositional representation of items in time requires conjunctive
representations of what $\times$ when (Eq.~\ref{eq:memgh}).  With particular
choices for temporal receptive fields (Eqs.~\ref{eq:h_F}~and~\ref{eq:gamma})
we find a good correspondence with a range of neural data (e.g.,
Figures~\ref{fig:Murray}~and~\ref{fig:Cueva}).  This raises the question of how
neural circuits could come to have receptive fields that obey these high-level
constraints.  One possibility is to simply have an RNN that obeys
Eq.~\ref{eq:RNNconst}, with temporal recurrent weights chosen to give
appropriate temporal receptive fields (see \citeA{LiuHowa20}).
However, a linear RNN would not be robust to perturbations.   To address this
question, we introduce a continuous attractor neural network (CANN) to
implement the Laplace Neural Manifold.  The neurons in this CANN will exhibit
conjunctive what $\times$ when receptive fields as specified by Eq.~\ref{eq:memgh}, with temporal receptive fields given by
Eqs.~\ref{eq:h_F}~and~\ref{eq:gamma} and logarithmic distribution of time
constants as given by Eq.~\ref{eq:log}.

If the temporal receptive fields across neurons differ by a single parameter
that can be mapped onto time, then it is possible to describe the time of a
past occurrence of a single stimulus by translating a pattern of activity along a
population.  As a thought experiment, imagine that we had built a CANN that
maintains a memory of the time of past events by constructing a bump of activity that moves at constant velocity along the population as a
function of time \cite{ZhanEtal22b}.  In this case, the activity of individual
neurons would rise and then fall as the bump approaches and then leaves their
location along the population.  The center of each neuron's  temporal
receptive field would depend only on its location along the population.  If
instead of a bump attractor, the network exhibited an edge across the
population that moved at constant velocity, then instead of circumscribed
temporal receptive fields, we would find monotonic temporal receptive fields.

\begin{figure*}[tb]
\begin{center}
\includegraphics[width=0.87\textwidth]{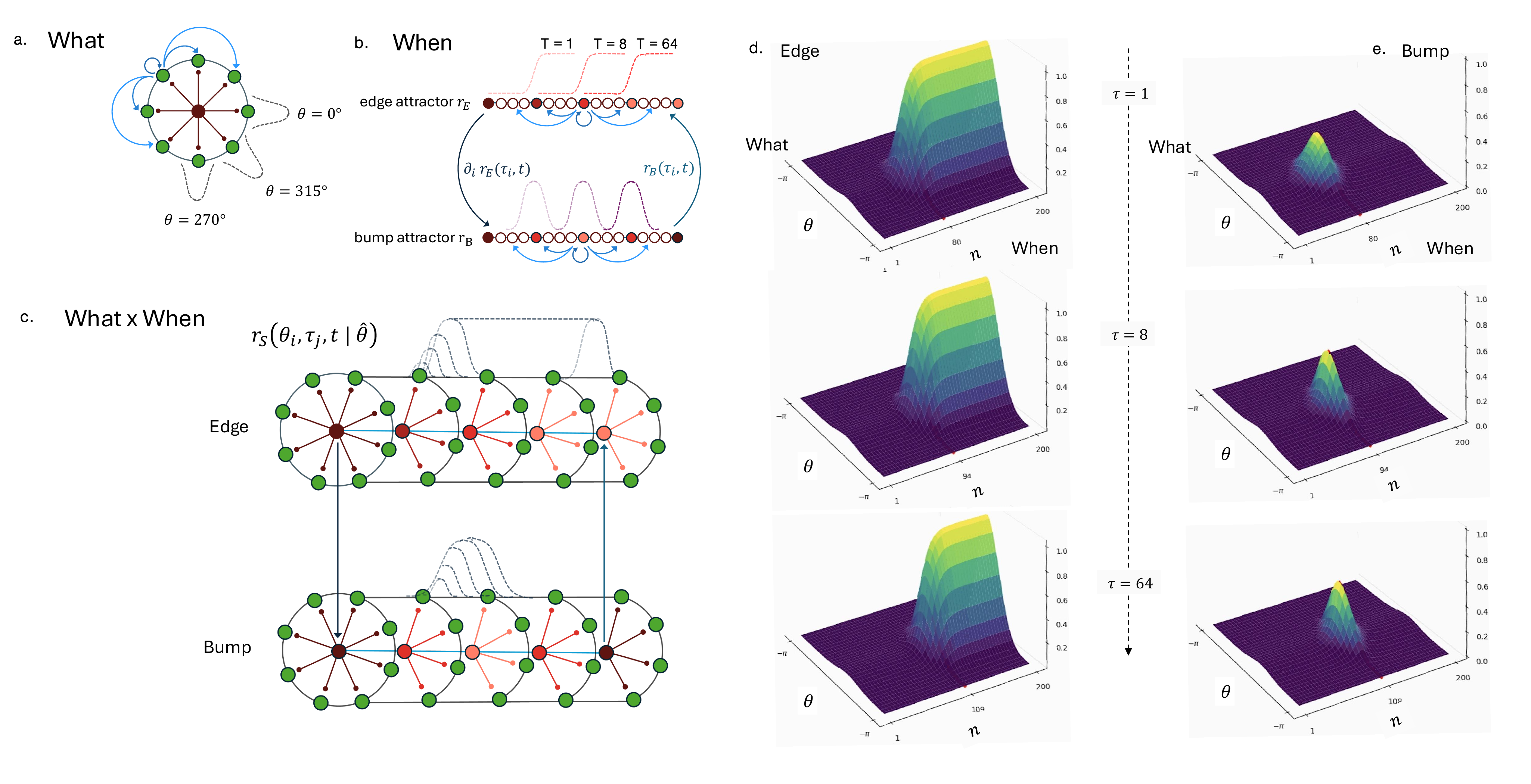}
\end{center}
\caption{ \emph{Continuous Attractor Neural Networks (CANNs) can be used to
	construct Laplace Neural Manifolds of what $\times$ when information.}
	 \textit{a.} 
     \textit{What:} A ring attractor maintains a persistent representation of the presented stimulus by sustaining a bump activation at that
	 angle.  Other kinds of stimulus information could be
	 maintained with appropriate circuits.  
	\textit{b.}
	 \textit{When:}
	A specialized set of line attractors can maintain an edge (Laplace)
	and a bump (Inverse Laplace) to implement temporal receptive fields
	evenly spaced over log time.  Recurrent connections
	maintain a particular shape of activity across the network; an edge
	for Laplace temporal receptive fields and a bump for Inverse Laplace
	temporal receptive fields.
	Connections between layers move the edge/bump at an appropriate speed,
	resulting in temporal receptive fields as a function of log time.
	\textit{c.}
	\textit{What $\times$ When:} A series of ring attractors
	as in \textit{a}, crossed with lines, as in \textit{b}  create a
	cylinder.  The  edge/bump attractor for time supplies global
	inhibition to the ring attractors coding for stimulus identity. 
	\textit{d, e.}
	Paired
	cylinders with backbones emulating the edge and bump respectively, can
	maintain the activity of Laplace and Inverse Laplace representations which shift as we progress along the remembered timeline $\tau$ akin to Fig. \ref{fig:LMNs},
	together forming a Laplace Neural Manifold for time.} \label{fig:CANN}
\end{figure*}

The temporal receptive fields described by
Eqs.~\ref{eq:h_F}~and~\ref{eq:gamma} differ from the receptive fields in our thought
experiment.  Let us return to a bump attractor
where the bump moves at a constant
rate.  In that case, the center of the temporal receptive fields would vary
systematically across neurons, but their width in time would be unaffected.
In contrast, in the temporal receptive fields used above
(Eqs.~\ref{eq:h_F}~and~\ref{eq:gamma}) the shape of the temporal receptive
fields depend only on $\tau/\tau_j$.  The temporal receptive fields of
different neurons are thus rescaled versions of one
another, rather than translated versions of one another.  Rescaling time
translates log time for the simple reason that $\log ax = \log a + \log x$. 
By choosing logarithmic time constants, Eq.~\ref{eq:log},  we can exploit the
fact that changing $\tau$ simply translates the representation of a single
event over the population (see Fig.~\ref{fig:LMNs}b,c, bottom).  For Laplace
neurons, the activity over the network as a function of $n$ takes the form of
an edge; for Inverse Laplace neurons, the activity takes the form of a bump.
In both cases, local connections favor nearby neurons to be in the same state.  
The bump network implementing the Inverse Laplace temporal receptive 
fields works as a standard local excitation/global inhibition CANN.
In the edge network, the two ends of the network
are clamped to be in an up and down states, so that an edge appears in between.  The edge can appear at any location 
such that the local connections are far from the clamped ends of the population. 
In contrast to the hypothetical CANN in which the bump moves at a constant
rate, Laplace Neural Manifolds require the edge/bump to move at a velocity
that decreases as $1/\tau$ \cite{DaniHowa25}.

In this section, we flesh out this idea to build a CANN that implements a
conjunctive working memory using a Laplace Neural Manifold focusing on the ODR task
from the previous section.  In this task, the monkey must remember the angle
of a visual stimulus, allowing a straightforward model for the representation of ``what'' using a ring attractor CANN (Fig.~\ref{fig:CANN}a), a topic that has been extensively studied for decades \cite{Amar77,Zhan96,RediEtal96,KimEtal17}. We
couple this to an edge/bump attractor to implement temporal receptive fields
(Fig.~\ref{fig:CANN}b).   In this model, paired CANNs with edge solutions and
bump solutions interact with one another.  Feedback between the two networks
causes the edge/bump complex to move at a velocity that goes down with the
location of the edge/bump, thus implementing logarithmically compressed
temporal receptive fields.  Coupling these two ideas together
(Fig.~\ref{fig:CANN}c) we can construct a single CANN that maintains
a compositional representation of what happened when.

In this model a stimulus, characterized by an angle describing its location
around the circle is presented at time zero.  This forms a bump of activity
along the `what' dimension of the cylinder, with an edge/bump at one edge of the
cylinder.  Feedback from the bump moves the edge from one moment to the next.
The strength of this connection decreases as a function of position along the
network, causing the edge/bump to move more slowly as time progresses
(Fig.~\ref{fig:CANN}d,e). 

\subsection{Separate CANNs for what and when information}
Before describing the combined network, for expository purposes, we first step
through the properties of the CANNs for what and when in isolation.

\subsubsection{Ring attractor tracks stimulus space `What'} 
To build a ring attractor to maintain information about the location of
the visual stimulus, we
assume that the neural interactions are Gaussian: $J(x,x') =
\frac{A}{\sqrt{2\pi a}}\exp\left[-\frac{(x-x')^2}{2a^2}\right]$, i.e., we
impose strong local excitation and weak global excitation over the ring
attractor network.
The input to each neuron  saturates under global activity-dependent
inhibition 
\begin{equation}
\sigma\left(\ringact(x);\ \inh, \rho\right) = \frac{\ringact(x, t)^2}{1 + \inh \rho
\|\ringact(\cdot, t)\|^2}
\end{equation}

given the inhibition strength $\inh$ and neural density
$\rho$. The dynamics of the synaptic input $\ringact(x,t)$ is governed by:
\begin{eqnarray} 
    \frac{d \ringact(x,t)}{dt} = -\ringact(x,t)+\rho \sum_{x'} J(x,x')
	\sigma\left(\ringact(x');\inh, \rho\right). 
\end{eqnarray}
We have chosen the time constant of the dynamics to be 1 and assume it
is much faster than the functional time constants of the receptive fields.
When the parameters are chosen properly the ring attractor network can
maintain a stable ``activity bump'' centered around the initial input $z$:
$\tilde{\ringact}(x|z) = \ringact_0 \exp\left[-\frac{(x-z)^2}{4a^2}\right]$, where $a$ is
the width of Gaussian interaction. 
The constant of proportionality $\ringact_0$ depends on the parameters of the
network, including the value of the inhibition parameter $\inh$.
This bump can shift around the ring in
response to changes in stimulus identity. The stimulus identity is thus
encoded by the manifold coordinate $\tilde{x}=z$.

\subsubsection{Paired edge-bump attractors track temporal space `When'}
Previous work has shown it is possible to build a continuous attractor model
for Laplace/inverse representations of a delta function
\cite{DaniHowa25} under the assumption that the time constants $\{\tau_n\}$ form
a geometric series. 
In this neural network the relative firing rate $\edgeact(\tau_i, t)$ of all neurons
is governed by the
dynamics:
\begin{multline*}
\frac{d \edgeact(\tau_i, t)}{dt}  = - \edgeact(\tau_i, t) + \sum_{j=1}^N \Wee(i,j) \phi(\edgeact; g) + I_{\text{ext}}^{\text{BE}}(\tau_i, t)     
\end{multline*}
where $\Wee$ is the recurrent matrix within the neural population, and
$\phi$ is the nonlinear transfer function $\phi(x) = \tanh (g x)$ with gain
$g$. Again, we assume the time constant of this dynamical system to be $1$ and
much less than the functional time constants $\tau_i$.  By carefully choosing the
recurrent matrix $\Wee$ and the gating $g$, the network shows monotonic exponential receptive
fields if we provide external input to clamp two edges of the network such
that at $i=1$ and $i=N$ $\edgeact{}$ are constrained to be  $-1$ and $1$
respectively.  With less careful tuning, any solution that gives an edge that moves
at the appropriate speed will give non-exponential monotonic receptive fields
\cite{DaniHowa25}.  

To cause the edge to move at a decreasing rate as a
function of time we provide a
dynamical input $I_{\text{ext}}^{\text{BE}}$ to the edge attractor. This input comes via another 
neuronal population  that forms a bump attractor in register with the edge attractor:
\begin{multline*}
    \frac{d \bumpact\left(\tau_i, t\right)}{dt} = - \bumpact(\tau_i, t) + \sum_{j=1}^N \Wbb(i,j) \phi(\bumpact; g) + I_{\text{ext}}^{EB}(\tau_i,t)
\end{multline*}
where $I_{\text{ext}}^{\text{EB}}(\tau_i,t)\equiv \edgeact(\tau_{i+1},t) -
\edgeact(\tau_{i},t)$ and $\Wbb$ is chosen order for the neural
network to maintain continuous bump-shaped states.  There is a natural one-to-one correspondence between the edge attractor and the bump attractor as the
corresponding units $i$ are associated with the same time constant $\tau_i$. 


The bump attractor receives input from the edge attractor
forming a bump at the position of the edge.  Meanwhile, the bump network provides input
to the edge attractor to stimulate it to move with a desired speed
$I_{\text{ext}}^{\text{BE}} (\tau_i, t) = v(\tau_i) \bumpact(\tau_i, t)$. By
customizing the speed function $v(t)$ the edge position grows logarithmically
in time:
\begin{eqnarray}
    \bar{n}(t) = \bar{n}_0+\log_a\frac{t}{t_0}
\end{eqnarray}
where $\bar{n}(t) = \arg\max_n \{\edgeact(\tau_{n+1},t)-\edgeact(\tau_n,t)\}$
denotes the edge position at time $t$, and $\bar{n}_0$, $t_0$ are free parameters
that determine the initial condition of the edge.
The edge/bump attractors exhibit dynamics that correspond to the
temporal receptive fields in Eqs.~\ref{eq:h_F}~and~\ref{eq:gamma} as
follows:
\begin{eqnarray}
	\frac{1}{2}\left[\edgeact(\tau_j, t)+1\right] &\propto&
	e^{-\frac{t}{\tau_j}},\nonumber\\
	\frac{1}{2}\left[\bumpact(\tau_j, t)+1\right] &\propto& \left(\frac{t}{\tau_j}\right)^{k} e^{-k(t/\tau_j)}.
\end{eqnarray}
Thus this edge/bump CANN can implement the temporal receptive fields
hypothesized for Laplace Neural Manifolds.

\subsection{Ring $\times$ edge-bump attractors track What $\times$ When}
The entire model is composed of a series of ring attractors that together form a 2-D
manifold that can be visualized as a cylinder. Each neuron in this 2-D
manifold is indexed by $i$ and $j$, indicating its preferred orientation $x_i$
on the ring and time constant $\tau_j$. There are connections between the
neurons in each ring of the cylinder, but in this simple model there are no
connections between the neurons across different rings. 
 The analytical solution of this continuous ring attractor shows that the
 height of the ``activity bump'' of  a particular ring depends on the inhibition strength $\inh$:
 \begin{eqnarray}
	 \ringact(x_i|\hat{x}) = \ringact_0(\kappa)
	 \exp\left[-\frac{(x_i-\bar{x})^2}{2a^2}\right]. 
\end{eqnarray}
To generate conjunctive what $\times$ when receptive fields, each cylinder is
controlled by an edge/bump 
``spine'' that controls the magnitude of the inhibition $\kappa$ of the ring
attractor for each value of $\tau_j$ at each moment.
To obtain conjunctive what $\times$ when receptive fields, the inhibition
strength at the ring at $\tau_j$ depends on the activation of the spine
edge-bump attractor, $\edgeact$/$\bumpact$.
The simulations in Fig.~\ref{fig:CANN}d~and ~\ref{fig:CANN}e use inhibition chosen as:
\begin{eqnarray}
\inh_{S}(\tau_j) = a_S \spineedge(\tau_j, t) + b_S \notag \\
\end{eqnarray}
with $S \in \{E,B\}$.
The parameter values in Fig.~\ref{fig:CANN}d~and~e
are $a_E=-9.60, b_E=10.71, a_B=-18.10, b_B=20.31$. These values were chosen such that
$\ringact_0(k)$ lies within $[0,1]$ for the edge attractor network as $\tau$ varies.

\subsubsection{Encoding of \emph{What} $\times$ \emph{When}}
Combining previous results, we are able to write out the dynamics of
Laplace/Inverse Laplace neural manifolds: 
\begin{align}
\frac{d \ringact_S(x_i, \tau_j, t|\hat{x})}{dt} &= -\ringact_S(x_i, \tau_j, t|\hat{x}) \nonumber \\
&\quad + \rho \sum_{k} J(x_i, x_k) \sigma\left(\ringact_S; \rho,
	\inh_{S}(\tau_j)\right).
\end{align}
Fig.~\ref{fig:CANN}d~and ~\ref{fig:CANN}e show results from this network choosing a
specific starting position $\hat{x}$.
Neurons in our 2-D manifold simultaneously encode both the `what' and `when'
information. The identity of the stimulus can
be readily decoded by observing the center of the bump across the rings, while
the timing of the stimulus can be inferred from the location of edge/bump 
along the `when' axis.  This simple circuit model is sufficient to generate
conjunctive temporal receptive fields with properties consistent with the
properties desired for $h_F$ and $h_{\f}$.

\section*{Discussion}

This paper explores the implications of a compositional neural representation
of what happened when in working memory.  Neurons in a population with such a
compositional representation should exhibit conjunctive receptive fields for
what and when, as has been reported in many brain regions
\cite{TigaEtal18a,TeraEtal17,TaxiEtal20}.   
This property makes it
straightforward to write out closed form expressions for the covariance matrix
of this population, which in turn allows us to work out the dynamics of the
population when studied using standard linear dimensionality reduction
techniques.  The low-dimensional dynamics depends dramatically on the choice of
temporal basis functions, even when the basis functions are related to one
another by a simple linear transformation.
With conjunctive receptive fields, the dimensionality of the space spanned by
the the population is controlled by the rank of the temporal covariance matrix,
allowing the density of basis functions to be directly assessed.  We show that a
logarithmic tiling of time, as proposed by work in cognitive psychology \cite{Fech60,Stev57,Murd60,LuceSupp02,BrowEtal07a,BalsGall09}
and supported by evidence from neuroscience \cite{CaoEtal22,GuoEtal21}
provides a reasonable approximation of empirical data.  Finally, we sketch out
a circuit model using continuous attractor neural networks that exhibits conjunctive
what $\times$ when receptive fields when a single item is present in working
memory.   

\subsection{Network Models in Neuroscience}

Early attractor models used persistent activity of neurons to encode a task
variable in working memory.  The current paper builds on and extends network models
that attempt to  reconcile stable task representations with
heterogeneous temporal dynamics \cite{DrucChkl12,MurrEtal17}. 
In those papers, the requirement on temporal dynamics is to leave some
stimulus-coding variable invariant to the passage of time.  
For instance, \citeA{DrucChkl12} required that the summed activity over
neurons coding for a particular stimulus is constant.  
The inverse Laplace cells (Eq.~\ref{eq:gamma}) in the current paper can satisfy that requirement
with appropriate normalization; because the temporal receptive fields decay
monotonically, the Laplace cells  (Eq.~\ref{eq:h_F}) do not satisfy this
requirement.

More broadly, the strategy in those papers is to keep the stimulus
representation within a ``null space" of the network dynamics. As long as the
changes in individual neuronal activity occur along directions in
the high-dimensional activity space that do not affect the decoded stimulus
representation, the stimulus can be  linearly decoded. 
For instance \citeA{MurrEtal17} used a
circuit model that has  a mnemonic coding subspace where the input stimulus
activates a stable representation, and a non-coding subspace which exhibits
temporal dynamics that are orthogonal to the coding subspace.

The major distinction between those previous models and the present paper is that
our approach is designed to construct a particular temporal representation
whereas previous papers did not place an emphasis on any particular form of
temporal dynamics.  
For instance \citeA{CuevEtal20} used a low-rank RNN \cite{BeirEtal23} (see
also \citeA{BeirEtal23}), to allow temporal dynamics along with a stable stimulus
representation.   The present approach starts
from the prior that memory for the passage of time requires a continuous
distribution of time constants, effectively tiling the time axis.  
If the time constants of a network must be trained by the requirements of a
particular task, this leaves the memory less able to express temporal
information in different tasks, or even  in situations where there is no
explicit demand for timing information.  For instance, we are aware of the
passage of time even when there is no explicit requirement to make a response
\cite{Huss66}.  Commitment to a timeline of what happened
when leaves the memory able to effectively decode \emph{any} what at
\emph{any} when.

\subsection{Attractor circuits as a mechanism for slow functional time constants}

The proposed circuit model using CANNs provides an existence proof that long
functional time constants need not be a consequence of intrinsic time
constants of individual neurons
\cite{FranEtal02,FranEtal06,LoewSomp03,TigaEtal15,MaroEtal25} nor of statistical
interactions between modes in a nearly-random RNN
\cite{DahmEtal19,helias2020statistical}. \citeA{SagoEtal24} described
conditions under which an RNN forms a line attractor (see also
\citeA{CanKris24,KrisEtal22}). Our results require further that line
attractors should be formed for all `what's that can exist in memory. In order
to get basis functions that are evenly spaced over log time, the network must
have degenerate eigenvalues in geometric series \emph{and} eigenvectors that
are translated versions of one another \cite{LiuHowa20}.  Different temporal
basis functions, which are apparently both observed in the mammalian brain,
also require distinct forms of line attractors. Thus the cognitive requirement
for compositional representations of what happened when provide strong
high-level constraints on physical models of working memory maintenance.

\subsection{Cognitive psychology of working memory}
In this paper we considered working memory for retention of a single item in continuous time.  Cognitive psychologists have studied
considerably more complex working memory tasks involving multiple stimuli that are to be remembered.  Time \emph{per se} has little effect on either visual working memory performance \cite{ShinEtal17,KahaSeku02}, nor on verbal working
memory performance \cite{BaddHitc77}.  However, there is widespread evidence that the amount of information that can be maintained in working memory in a highly veridical manner is finite.
For many decades the dominant view has been that working memory maintains a discrete number of items in a discrete number of ``slots''
\cite{AtkiShif68,BaddHitc74,LuckVoge97}.  
However, more recently, cognitive psychologists' understanding of the nature of working memory has become much more subtle
\cite{MaEtal14,BradEtal16,SchuEtal20}.  In the words of \citeA{MaEtal14}
``working memory might better be conceptualized as a limited resource that is distributed flexibly among all items to be maintained in memory.''

It has long been understood that recurrent attractor neural networks provide a natural means to understand capacity limitations \cite{Gros69,Gros78}. Mutual inhibition controls the number of attractors, or bumps, that can be simultaneously sustained by recurrent activation.   One can readily imagine extending the model described in Fig.~\ref{fig:CANN} to allow multiple bumps to coexist. Given a particular structure of the ``what'' network, depending on how
inhibition flows across the network, one can control capacity within a time point (as is typically done in visual working memory experiments) or across time points (as is typically done in verbal working memory experiments. In light of recent experimental work, much care should be taken in allowing stimulus features to cooperate and compete in working memory.

The wiring requirements necessary to extend this approach such that \emph{all possible stimuli} can be maintained in working memory through continuous time seem daunting.  For instance,  what if the task also required memory for the color of the stimuli as well as their location?  Or the orientation and color of a bar at an angual location?  Or the identity of a letter at an angular location?  The circuit would have to either be extremely complicated \emph{a priori} or be able to be dynamically configured for task-relevant information.

Perhaps recent work on gated RNNs provides a way for these continuous
attractor networks to dynamically form in response to task demands
\cite{KrisEtal22,CanKris24}. In this way, gated input would dynamically
specify the features that can be maintained in a continuous attractor network for what information. 

\subsection{Time and space and number}

 One could have easily generalized the compositionality argument for what
$\times$ when to what $\times$ spatial position.  Indeed,
\citeA{Weyl22} used precisely the same argument for space in developing an axiomatic development of space-time in general relativity. The requirement of compositional representations for what and when leads naturally to conjunctive neural codes.

In addition to physical space, one might have made analogous arguments for any number of continuous variables.
Conjunctive codes for what $\times$ when is thus a special case of mixed
selectivity, which has been proposed as a general property for neural codes \cite{RigoEtal13} and has been observed for many different variables in neural
representations 
\cite{MantEtal13, WardEtal10,JohnEtal20,DangEtal21,WallEtal22}. Conjunctive receptive fields are also closely related to gain fields
which were proposed for coordinate transformations and efficient function
learning \cite{PougSejn97,SaliAbbo01}.

Numerosity is a variable that can be composed (in principle) with any kind of
object.  Mammals \cite{GallGelm92,DehaBran11} and other animals
\cite{KirsNied23} are equipped with a number sense.
This number sense appears to be computed over a neural  scale that maps
receptive fields onto the log of numerosity
\cite{NiedMill03,Deha03,NiedDeha09}, presumably using a set of basis functions
distributed over a logarithmic number line.

\subsection{Theoretically-driven neural data analysis techniques}
The appearance of population dynamics projected onto a low-dimensional PCA
space changed dramatically depending on how we chose temporal basis functions
(Fig.~\ref{fig:Murray}). 
Even though the temporal basis functions are simply related by a linear
transformation, one might have concluded that they reflect very different
coding schemes if one only looked at the PCA results. 
For instance, one may have concluded that the Laplace representation shows a
stable subspace with persistent firing \cite{MurrEtal17,ConsEtal18}
whereas the inverse space shows sustained delay-period coding without
persistent firing \cite{KingDeha14,LundEtal18a}.

This is another example that extreme caution should be exercised in using linear dimensionality reduction techniques in neuroscience \cite{LebeEtal19,DeChau23,Shin23}.
\citeA{JazaOsto21} write that ``\ldots without concrete computational hypotheses, it could be extremely challenging to interpret measures of dimensionality.''   If neural codes use basis functions tiling continuous variables out in the world, then the linear dimensionality of the neural code is in principle unbounded \cite{ManlEtal24}.  Depending on the choice of basis functions, the principle components of the population may, or may not be readily interpretable, even if we know \emph{a priori} the continuous variables that are being coded by the population.

A number of well-studied non-linear dimensionality reduction techniques exist 
\cite{BelkNiyo03,KohlEtal21,TeneEtal00,RoweSaul00,McInEtal18,VanHint08}.
Although still vastly outnumbered by data analyses using linear dimensionality
reduction techniques, some neuroscience work has made use of non-linear
dimensionality reduction \cite{ChauEtal19,NiehEtal21}  
(see \citeNP{LangEtal23} for a review).
Resolving the empirical question of the generality of compositional codes
using conjunctive basis functions will require careful experimentation and
development of new data analysis tools.

Given a hypothesis about relevant variables, it should be possible to
distinguish if the decomposition of the covariance matrix as asserted in
Eq.~\ref{eq:covproduct} is valid.  This would establish that a
conjunctive code exists.  Proximity of the parameters describing the receptive
fields, e.g., $s_i$, $\tau^\star_i$, $n$, provide a natural metric for
proximity along the population manifold.  However, these parameters need to be
recovered from the data, preferably without a strong prior on the form of the
functions describing the receptive fields.

\acknowledgements{
This work was supported by a Rajen Kilachand Fund award to Michael Hasselmo and NIMH
R01MH132171 to MWH. 
}

\bibliography{thesis,bibdesk}

\begin{thebibliography}{}

\bibitem [\protect \citeauthoryear {%
Akhlaghpour%
\ \protect \BOthers {.}}{%
Akhlaghpour%
\ \protect \BOthers {.}}{%
{\protect \APACyear {2016}}%
}]{%
AkhlEtal16}
\APACinsertmetastar {%
AkhlEtal16}%
\begin{APACrefauthors}%
Akhlaghpour, H.%
, Wiskerke, J.%
, Choi, J\BPBI Y.%
, Taliaferro, J\BPBI P.%
, Au, J.%
\BCBL {}\ \BBA {} Witten, I.%
\end{APACrefauthors}%
\unskip\
\newblock
\APACrefYearMonthDay{2016}{}{}.
\newblock
{\BBOQ}\APACrefatitle {Dissociated sequential activity and stimulus encoding in the dorsomedial striatum during spatial working memory} {Dissociated sequential activity and stimulus encoding in the dorsomedial striatum during spatial working memory}.{\BBCQ}
\newblock
\APACjournalVolNumPages{eLife}{5}{}{e19507}.
\PrintBackRefs{\CurrentBib}

\bibitem [\protect \citeauthoryear {%
Amari%
}{%
Amari%
}{%
{\protect \APACyear {1977}}%
}]{%
Amar77}
\APACinsertmetastar {%
Amar77}%
\begin{APACrefauthors}%
Amari, S.%
\end{APACrefauthors}%
\unskip\
\newblock
\APACrefYearMonthDay{1977}{}{}.
\newblock
{\BBOQ}\APACrefatitle {Dynamics of pattern formation in lateral-inhibition type neural fields} {Dynamics of pattern formation in lateral-inhibition type neural fields}.{\BBCQ}
\newblock
\APACjournalVolNumPages{Biological cybernetics}{27}{2}{77--87}.
\PrintBackRefs{\CurrentBib}

\bibitem [\protect \citeauthoryear {%
Atanas%
\ \protect \BOthers {.}}{%
Atanas%
\ \protect \BOthers {.}}{%
{\protect \APACyear {2023}}%
}]{%
AtanEtal23}
\APACinsertmetastar {%
AtanEtal23}%
\begin{APACrefauthors}%
Atanas, A\BPBI A.%
, Kim, J.%
, Wang, Z.%
, Bueno, E.%
, Becker, M.%
, Kang, D.%
\BDBL {}others%
\end{APACrefauthors}%
\unskip\
\newblock
\APACrefYearMonthDay{2023}{}{}.
\newblock
{\BBOQ}\APACrefatitle {Brain-wide representations of behavior spanning multiple timescales and states in C. elegans} {Brain-wide representations of behavior spanning multiple timescales and states in c. elegans}.{\BBCQ}
\newblock
\APACjournalVolNumPages{Cell}{186}{19}{4134--4151}.
\PrintBackRefs{\CurrentBib}

\bibitem [\protect \citeauthoryear {%
Atkinson%
\ \BBA {} Shiffrin%
}{%
Atkinson%
\ \BBA {} Shiffrin%
}{%
{\protect \APACyear {1968}}%
}]{%
AtkiShif68}
\APACinsertmetastar {%
AtkiShif68}%
\begin{APACrefauthors}%
Atkinson, R\BPBI C.%
\BCBT {}\ \BBA {} Shiffrin, R\BPBI M.%
\end{APACrefauthors}%
\unskip\
\newblock
\APACrefYearMonthDay{1968}{}{}.
\newblock
{\BBOQ}\APACrefatitle {Human memory: A proposed system and its control processes} {Human memory: A proposed system and its control processes}.{\BBCQ}
\newblock
\BIn{} K\BPBI W.~Spence\ \BBA {} J\BPBI T.~Spence\ (\BEDS), \APACrefbtitle {The Psychology of Learning and Motivation} {The psychology of learning and motivation}\ (\BVOL~2, \BPG~89-105).
\newblock
\APACaddressPublisher{New York}{Academic Press}.
\PrintBackRefs{\CurrentBib}

\bibitem [\protect \citeauthoryear {%
Baddeley%
\ \BBA {} Hitch%
}{%
Baddeley%
\ \BBA {} Hitch%
}{%
{\protect \APACyear {1974}}%
}]{%
BaddHitc74}
\APACinsertmetastar {%
BaddHitc74}%
\begin{APACrefauthors}%
Baddeley, A\BPBI D.%
\BCBT {}\ \BBA {} Hitch, G\BPBI J.%
\end{APACrefauthors}%
\unskip\
\newblock
\APACrefYearMonthDay{1974}{}{}.
\newblock
{\BBOQ}\APACrefatitle {Working memory} {Working memory}.{\BBCQ}
\newblock
\BIn{} G\BPBI H.~Bower\ (\BED), \APACrefbtitle {The psychology of learning and motivation: Advances in research and theory} {The psychology of learning and motivation: Advances in research and theory}\ (\BVOL~8, \BPG~47-90).
\newblock
\APACaddressPublisher{New York}{Academic Press}.
\PrintBackRefs{\CurrentBib}

\bibitem [\protect \citeauthoryear {%
Baddeley%
\ \BBA {} Hitch%
}{%
Baddeley%
\ \BBA {} Hitch%
}{%
{\protect \APACyear {1977}}%
}]{%
BaddHitc77}
\APACinsertmetastar {%
BaddHitc77}%
\begin{APACrefauthors}%
Baddeley, A\BPBI D.%
\BCBT {}\ \BBA {} Hitch, G\BPBI J.%
\end{APACrefauthors}%
\unskip\
\newblock
\APACrefYearMonthDay{1977}{}{}.
\newblock
{\BBOQ}\APACrefatitle {Recency reexamined} {Recency reexamined}.{\BBCQ}
\newblock
\BIn{} S.~Dornic\ (\BED), \APACrefbtitle {Attention and Performance {VI}} {Attention and performance {VI}}\ (\BPG~647-667).
\newblock
\APACaddressPublisher{Hillsdale, NJ}{Erlbaum}.
\PrintBackRefs{\CurrentBib}

\bibitem [\protect \citeauthoryear {%
Balsam%
\ \BBA {} Gallistel%
}{%
Balsam%
\ \BBA {} Gallistel%
}{%
{\protect \APACyear {2009}}%
}]{%
BalsGall09}
\APACinsertmetastar {%
BalsGall09}%
\begin{APACrefauthors}%
Balsam, P\BPBI D.%
\BCBT {}\ \BBA {} Gallistel, C\BPBI R.%
\end{APACrefauthors}%
\unskip\
\newblock
\APACrefYearMonthDay{2009}{}{}.
\newblock
{\BBOQ}\APACrefatitle {Temporal maps and informativeness in associative learning.} {Temporal maps and informativeness in associative learning.}{\BBCQ}
\newblock
\APACjournalVolNumPages{Trends in Neuroscience}{32}{2}{73--78}.
\PrintBackRefs{\CurrentBib}

\bibitem [\protect \citeauthoryear {%
Beiran%
, Meirhaeghe%
, Sohn%
, Jazayeri%
\BCBL {}\ \BBA {} Ostojic%
}{%
Beiran%
\ \protect \BOthers {.}}{%
{\protect \APACyear {2023}}%
}]{%
BeirEtal23}
\APACinsertmetastar {%
BeirEtal23}%
\begin{APACrefauthors}%
Beiran, M.%
, Meirhaeghe, N.%
, Sohn, H.%
, Jazayeri, M.%
\BCBL {}\ \BBA {} Ostojic, S.%
\end{APACrefauthors}%
\unskip\
\newblock
\APACrefYearMonthDay{2023}{}{}.
\newblock
{\BBOQ}\APACrefatitle {Parametric control of flexible timing through low-dimensional neural manifolds} {Parametric control of flexible timing through low-dimensional neural manifolds}.{\BBCQ}
\newblock
\APACjournalVolNumPages{Neuron}{}{}{}.
\PrintBackRefs{\CurrentBib}

\bibitem [\protect \citeauthoryear {%
Belkin%
\ \BBA {} Niyogi%
}{%
Belkin%
\ \BBA {} Niyogi%
}{%
{\protect \APACyear {2003}}%
}]{%
BelkNiyo03}
\APACinsertmetastar {%
BelkNiyo03}%
\begin{APACrefauthors}%
Belkin, M.%
\BCBT {}\ \BBA {} Niyogi, P.%
\end{APACrefauthors}%
\unskip\
\newblock
\APACrefYearMonthDay{2003}{}{}.
\newblock
{\BBOQ}\APACrefatitle {Laplacian eigenmaps for dimensionality reduction and data representation} {Laplacian eigenmaps for dimensionality reduction and data representation}.{\BBCQ}
\newblock
\APACjournalVolNumPages{Neural Computation}{15}{6}{1373--1396}.
\PrintBackRefs{\CurrentBib}

\bibitem [\protect \citeauthoryear {%
Bergson%
}{%
Bergson%
}{%
{\protect \APACyear {1910}}%
}]{%
Berg10}
\APACinsertmetastar {%
Berg10}%
\begin{APACrefauthors}%
Bergson, H.%
\end{APACrefauthors}%
\unskip\
\newblock
\APACrefYear{1910}.
\newblock
\APACrefbtitle {Time and Free Will: An Essay on the Immediate Data of Consciousness} {Time and free will: An essay on the immediate data of consciousness}.
\newblock
\APACaddressPublisher{}{G. Allen \& Unwin}.
\PrintBackRefs{\CurrentBib}

\bibitem [\protect \citeauthoryear {%
Bernardi%
\ \protect \BOthers {.}}{%
Bernardi%
\ \protect \BOthers {.}}{%
{\protect \APACyear {2020}}%
}]{%
BernEtal20}
\APACinsertmetastar {%
BernEtal20}%
\begin{APACrefauthors}%
Bernardi, S.%
, Benna, M\BPBI K.%
, Rigotti, M.%
, Munuera, J.%
, Fusi, S.%
\BCBL {}\ \BBA {} Salzman, C\BPBI D.%
\end{APACrefauthors}%
\unskip\
\newblock
\APACrefYearMonthDay{2020}{}{}.
\newblock
{\BBOQ}\APACrefatitle {The geometry of abstraction in the hippocampus and prefrontal cortex} {The geometry of abstraction in the hippocampus and prefrontal cortex}.{\BBCQ}
\newblock
\APACjournalVolNumPages{Cell}{183}{4}{954--967}.
\PrintBackRefs{\CurrentBib}

\bibitem [\protect \citeauthoryear {%
Brady%
, St{\"o}rmer%
\BCBL {}\ \BBA {} Alvarez%
}{%
Brady%
\ \protect \BOthers {.}}{%
{\protect \APACyear {2016}}%
}]{%
BradEtal16}
\APACinsertmetastar {%
BradEtal16}%
\begin{APACrefauthors}%
Brady, T\BPBI F.%
, St{\"o}rmer, V\BPBI S.%
\BCBL {}\ \BBA {} Alvarez, G\BPBI A.%
\end{APACrefauthors}%
\unskip\
\newblock
\APACrefYearMonthDay{2016}{}{}.
\newblock
{\BBOQ}\APACrefatitle {Working memory is not fixed-capacity: More active storage capacity for real-world objects than for simple stimuli} {Working memory is not fixed-capacity: More active storage capacity for real-world objects than for simple stimuli}.{\BBCQ}
\newblock
\APACjournalVolNumPages{Proceedings of the National Academy of Sciences}{113}{27}{7459--7464}.
\PrintBackRefs{\CurrentBib}

\bibitem [\protect \citeauthoryear {%
Bright%
\ \protect \BOthers {.}}{%
Bright%
\ \protect \BOthers {.}}{%
{\protect \APACyear {2020}}%
}]{%
BrigEtal20}
\APACinsertmetastar {%
BrigEtal20}%
\begin{APACrefauthors}%
Bright, I\BPBI M.%
, Meister, M\BPBI L\BPBI R.%
, Cruzado, N\BPBI A.%
, Tiganj, Z.%
, Buffalo, E\BPBI A.%
\BCBL {}\ \BBA {} Howard, M\BPBI W.%
\end{APACrefauthors}%
\unskip\
\newblock
\APACrefYearMonthDay{2020}{}{}.
\newblock
{\BBOQ}\APACrefatitle {A temporal record of the past with a spectrum of time constants in the monkey entorhinal cortex} {A temporal record of the past with a spectrum of time constants in the monkey entorhinal cortex}.{\BBCQ}
\newblock
\APACjournalVolNumPages{Proceedings of the National Academy of Sciences}{117}{}{20274-20283}.
\PrintBackRefs{\CurrentBib}

\bibitem [\protect \citeauthoryear {%
Brown%
, Neath%
\BCBL {}\ \BBA {} Chater%
}{%
Brown%
\ \protect \BOthers {.}}{%
{\protect \APACyear {2007}}%
}]{%
BrowEtal07a}
\APACinsertmetastar {%
BrowEtal07a}%
\begin{APACrefauthors}%
Brown, G\BPBI D\BPBI A.%
, Neath, I.%
\BCBL {}\ \BBA {} Chater, N.%
\end{APACrefauthors}%
\unskip\
\newblock
\APACrefYearMonthDay{2007}{}{}.
\newblock
{\BBOQ}\APACrefatitle {A temporal ratio model of memory.} {A temporal ratio model of memory.}{\BBCQ}
\newblock
\APACjournalVolNumPages{Psychological Review}{114}{3}{539-76}.
\PrintBackRefs{\CurrentBib}

\bibitem [\protect \citeauthoryear {%
Can%
\ \BBA {} Krishnamurthy%
}{%
Can%
\ \BBA {} Krishnamurthy%
}{%
{\protect \APACyear {2024}}%
}]{%
CanKris24}
\APACinsertmetastar {%
CanKris24}%
\begin{APACrefauthors}%
Can, T.%
\BCBT {}\ \BBA {} Krishnamurthy, K.%
\end{APACrefauthors}%
\unskip\
\newblock
\APACrefYearMonthDay{2024}{}{}.
\newblock
{\BBOQ}\APACrefatitle {Emergence of memory manifolds} {Emergence of memory manifolds}.{\BBCQ}
\newblock
\APACjournalVolNumPages{arXiv preprint arXiv:2109.03879}{}{}{}.
\PrintBackRefs{\CurrentBib}

\bibitem [\protect \citeauthoryear {%
Cao%
, Bladon%
, Charczynski%
, Hasselmo%
\BCBL {}\ \BBA {} Howard%
}{%
Cao%
\ \protect \BOthers {.}}{%
{\protect \APACyear {2022}}%
}]{%
CaoEtal22}
\APACinsertmetastar {%
CaoEtal22}%
\begin{APACrefauthors}%
Cao, R.%
, Bladon, J\BPBI H.%
, Charczynski, S\BPBI J.%
, Hasselmo, M.%
\BCBL {}\ \BBA {} Howard, M.%
\end{APACrefauthors}%
\unskip\
\newblock
\APACrefYearMonthDay{2022}{}{}.
\newblock
{\BBOQ}\APACrefatitle {Internally generated time in the rodent hippocampus is logarithmically compressed} {Internally generated time in the rodent hippocampus is logarithmically compressed}.{\BBCQ}
\newblock
\APACjournalVolNumPages{eLife}{https://doi.org/10.7554/eLife.75353}{}{}.
\PrintBackRefs{\CurrentBib}

\bibitem [\protect \citeauthoryear {%
Cao%
, Bright%
\BCBL {}\ \BBA {} Howard%
}{%
Cao%
\ \protect \BOthers {.}}{%
{\protect \APACyear {2024}}%
}]{%
CaoEtal24}
\APACinsertmetastar {%
CaoEtal24}%
\begin{APACrefauthors}%
Cao, R.%
, Bright, I\BPBI M.%
\BCBL {}\ \BBA {} Howard, M\BPBI W.%
\end{APACrefauthors}%
\unskip\
\newblock
\APACrefYearMonthDay{2024}{}{}.
\newblock
{\BBOQ}\APACrefatitle {Ramping cells in the rodent medial prefrontal cortex encode time to past and future events via real Laplace transform} {Ramping cells in the rodent medial prefrontal cortex encode time to past and future events via real laplace transform}.{\BBCQ}
\newblock
\APACjournalVolNumPages{Proceedings of the National Academy of Sciences}{121}{38}{e2404169121}.
\PrintBackRefs{\CurrentBib}

\bibitem [\protect \citeauthoryear {%
Chaudhuri%
, Ger{\c{c}}ek%
, Pandey%
, Peyrache%
\BCBL {}\ \BBA {} Fiete%
}{%
Chaudhuri%
\ \protect \BOthers {.}}{%
{\protect \APACyear {2019}}%
}]{%
ChauEtal19}
\APACinsertmetastar {%
ChauEtal19}%
\begin{APACrefauthors}%
Chaudhuri, R.%
, Ger{\c{c}}ek, B.%
, Pandey, B.%
, Peyrache, A.%
\BCBL {}\ \BBA {} Fiete, I.%
\end{APACrefauthors}%
\unskip\
\newblock
\APACrefYearMonthDay{2019}{}{}.
\newblock
{\BBOQ}\APACrefatitle {The intrinsic attractor manifold and population dynamics of a canonical cognitive circuit across waking and sleep} {The intrinsic attractor manifold and population dynamics of a canonical cognitive circuit across waking and sleep}.{\BBCQ}
\newblock
\APACjournalVolNumPages{Nature neuroscience}{22}{9}{1512--1520}.
\PrintBackRefs{\CurrentBib}

\bibitem [\protect \citeauthoryear {%
Constantinidis%
\ \protect \BOthers {.}}{%
Constantinidis%
\ \protect \BOthers {.}}{%
{\protect \APACyear {2018}}%
}]{%
ConsEtal18}
\APACinsertmetastar {%
ConsEtal18}%
\begin{APACrefauthors}%
Constantinidis, C.%
, Funahashi, S.%
, Lee, D.%
, Murray, J\BPBI D.%
, Qi, X\BHBI L.%
, Wang, M.%
\BCBL {}\ \BBA {} Arnsten, A\BPBI F\BPBI T.%
\end{APACrefauthors}%
\unskip\
\newblock
\APACrefYearMonthDay{2018}{}{}.
\newblock
{\BBOQ}\APACrefatitle {Persistent Spiking Activity Underlies Working Memory} {Persistent spiking activity underlies working memory}.{\BBCQ}
\newblock
\APACjournalVolNumPages{Journal of Neuroscience}{38}{32}{7020-7028}.
\newblock
\begin{APACrefDOI} \doi{10.1523/JNEUROSCI.2486-17.2018} \end{APACrefDOI}
\PrintBackRefs{\CurrentBib}

\bibitem [\protect \citeauthoryear {%
Cueva%
\ \protect \BOthers {.}}{%
Cueva%
\ \protect \BOthers {.}}{%
{\protect \APACyear {2020}}%
}]{%
CuevEtal20}
\APACinsertmetastar {%
CuevEtal20}%
\begin{APACrefauthors}%
Cueva, C\BPBI J.%
, Saez, A.%
, Marcos, E.%
, Genovesio, A.%
, Jazayeri, M.%
, Romo, R.%
\BDBL {}Fusi, S.%
\end{APACrefauthors}%
\unskip\
\newblock
\APACrefYearMonthDay{2020}{}{}.
\newblock
{\BBOQ}\APACrefatitle {Low-dimensional dynamics for working memory and time encoding} {Low-dimensional dynamics for working memory and time encoding}.{\BBCQ}
\newblock
\APACjournalVolNumPages{Proceedings of the National Academy of Sciences}{117}{37}{23021--23032}.
\PrintBackRefs{\CurrentBib}

\bibitem [\protect \citeauthoryear {%
Dahmen%
, Gr{\"u}n%
, Diesmann%
\BCBL {}\ \BBA {} Helias%
}{%
Dahmen%
\ \protect \BOthers {.}}{%
{\protect \APACyear {2019}}%
}]{%
DahmEtal19}
\APACinsertmetastar {%
DahmEtal19}%
\begin{APACrefauthors}%
Dahmen, D.%
, Gr{\"u}n, S.%
, Diesmann, M.%
\BCBL {}\ \BBA {} Helias, M.%
\end{APACrefauthors}%
\unskip\
\newblock
\APACrefYearMonthDay{2019}{}{}.
\newblock
{\BBOQ}\APACrefatitle {Second type of criticality in the brain uncovers rich multiple-neuron dynamics} {Second type of criticality in the brain uncovers rich multiple-neuron dynamics}.{\BBCQ}
\newblock
\APACjournalVolNumPages{Proceedings of the National Academy of Sciences}{116}{26}{13051--13060}.
\PrintBackRefs{\CurrentBib}

\bibitem [\protect \citeauthoryear {%
Dang%
, Jaffe%
, Qi%
\BCBL {}\ \BBA {} Constantinidis%
}{%
Dang%
\ \protect \BOthers {.}}{%
{\protect \APACyear {2021}}%
}]{%
DangEtal21}
\APACinsertmetastar {%
DangEtal21}%
\begin{APACrefauthors}%
Dang, W.%
, Jaffe, R\BPBI J.%
, Qi, X\BHBI L.%
\BCBL {}\ \BBA {} Constantinidis, C.%
\end{APACrefauthors}%
\unskip\
\newblock
\APACrefYearMonthDay{2021}{}{}.
\newblock
{\BBOQ}\APACrefatitle {Emergence of nonlinear mixed selectivity in prefrontal cortex after training} {Emergence of nonlinear mixed selectivity in prefrontal cortex after training}.{\BBCQ}
\newblock
\APACjournalVolNumPages{Journal of Neuroscience}{41}{35}{7420--7434}.
\PrintBackRefs{\CurrentBib}

\bibitem [\protect \citeauthoryear {%
Daniels%
\ \BBA {} Howard%
}{%
Daniels%
\ \BBA {} Howard%
}{%
{\protect \APACyear {2025}}%
}]{%
DaniHowa25}
\APACinsertmetastar {%
DaniHowa25}%
\begin{APACrefauthors}%
Daniels, B\BPBI C.%
\BCBT {}\ \BBA {} Howard, M\BPBI W.%
\end{APACrefauthors}%
\unskip\
\newblock
\APACrefYearMonthDay{2025}{}{}.
\newblock
{\BBOQ}\APACrefatitle {Continuous Attractor Networks for Laplace Neural Manifolds} {Continuous attractor networks for laplace neural manifolds}.{\BBCQ}
\newblock
\APACjournalVolNumPages{Computational Brain \& Behavior}{}{}{1--18}.
\PrintBackRefs{\CurrentBib}

\bibitem [\protect \citeauthoryear {%
De%
\ \BBA {} Chaudhuri%
}{%
De%
\ \BBA {} Chaudhuri%
}{%
{\protect \APACyear {2023}}%
}]{%
DeChau23}
\APACinsertmetastar {%
DeChau23}%
\begin{APACrefauthors}%
De, A.%
\BCBT {}\ \BBA {} Chaudhuri, R.%
\end{APACrefauthors}%
\unskip\
\newblock
\APACrefYearMonthDay{2023}{}{}.
\newblock
{\BBOQ}\APACrefatitle {Common population codes produce extremely nonlinear neural manifolds} {Common population codes produce extremely nonlinear neural manifolds}.{\BBCQ}
\newblock
\APACjournalVolNumPages{Proceedings of the National Academy of Sciences}{120}{39}{e2305853120}.
\PrintBackRefs{\CurrentBib}

\bibitem [\protect \citeauthoryear {%
Dehaene%
}{%
Dehaene%
}{%
{\protect \APACyear {2003}}%
}]{%
Deha03}
\APACinsertmetastar {%
Deha03}%
\begin{APACrefauthors}%
Dehaene, S.%
\end{APACrefauthors}%
\unskip\
\newblock
\APACrefYearMonthDay{2003}{}{}.
\newblock
{\BBOQ}\APACrefatitle {The neural basis of the {Weber-Fechner} law: a logarithmic mental number line} {The neural basis of the {Weber-Fechner} law: a logarithmic mental number line}.{\BBCQ}
\newblock
\APACjournalVolNumPages{Trends in Cognitive Sciences}{7}{4}{145-147}.
\PrintBackRefs{\CurrentBib}

\bibitem [\protect \citeauthoryear {%
Dehaene%
\ \BBA {} Brannon%
}{%
Dehaene%
\ \BBA {} Brannon%
}{%
{\protect \APACyear {2011}}%
}]{%
DehaBran11}
\APACinsertmetastar {%
DehaBran11}%
\begin{APACrefauthors}%
Dehaene, S.%
\BCBT {}\ \BBA {} Brannon, E.%
\end{APACrefauthors}%
\unskip\
\newblock
\APACrefYear{2011}.
\newblock
\APACrefbtitle {Space, time and number in the brain: Searching for the foundations of mathematical thought} {Space, time and number in the brain: Searching for the foundations of mathematical thought}.
\newblock
\APACaddressPublisher{}{Academic Press}.
\PrintBackRefs{\CurrentBib}

\bibitem [\protect \citeauthoryear {%
De~Vries%
\ \BBA {} Principe%
}{%
De~Vries%
\ \BBA {} Principe%
}{%
{\protect \APACyear {1992}}%
}]{%
deVrPrin92}
\APACinsertmetastar {%
deVrPrin92}%
\begin{APACrefauthors}%
De~Vries, B.%
\BCBT {}\ \BBA {} Principe, J\BPBI C.%
\end{APACrefauthors}%
\unskip\
\newblock
\APACrefYearMonthDay{1992}{}{}.
\newblock
{\BBOQ}\APACrefatitle {The gamma model---A new neural model for temporal processing} {The gamma model---a new neural model for temporal processing}.{\BBCQ}
\newblock
\APACjournalVolNumPages{Neural networks}{5}{4}{565--576}.
\PrintBackRefs{\CurrentBib}

\bibitem [\protect \citeauthoryear {%
Druckmann%
\ \BBA {} Chklovskii%
}{%
Druckmann%
\ \BBA {} Chklovskii%
}{%
{\protect \APACyear {2012}}%
}]{%
DrucChkl12}
\APACinsertmetastar {%
DrucChkl12}%
\begin{APACrefauthors}%
Druckmann, S.%
\BCBT {}\ \BBA {} Chklovskii, D\BPBI B.%
\end{APACrefauthors}%
\unskip\
\newblock
\APACrefYearMonthDay{2012}{}{}.
\newblock
{\BBOQ}\APACrefatitle {Neuronal circuits underlying persistent representations despite time varying activity} {Neuronal circuits underlying persistent representations despite time varying activity}.{\BBCQ}
\newblock
\APACjournalVolNumPages{Current Biology}{22}{22}{2095--2103}.
\PrintBackRefs{\CurrentBib}

\bibitem [\protect \citeauthoryear {%
Fechner%
}{%
Fechner%
}{%
{\protect \APACyear {1860/1912}}%
}]{%
Fech60}
\APACinsertmetastar {%
Fech60}%
\begin{APACrefauthors}%
Fechner, G.%
\end{APACrefauthors}%
\unskip\
\newblock
\APACrefYear{1860/1912}.
\newblock
\APACrefbtitle {Elements of Psychophysics. {Vol. I.}} {Elements of psychophysics. {Vol. I.}}
\newblock
\APACaddressPublisher{}{Houghton Mifflin}.
\PrintBackRefs{\CurrentBib}

\bibitem [\protect \citeauthoryear {%
Frans\'{e}n%
, Alonso%
\BCBL {}\ \BBA {} Hasselmo%
}{%
Frans\'{e}n%
\ \protect \BOthers {.}}{%
{\protect \APACyear {2002}}%
}]{%
FranEtal02}
\APACinsertmetastar {%
FranEtal02}%
\begin{APACrefauthors}%
Frans\'{e}n, E.%
, Alonso, A\BPBI A.%
\BCBL {}\ \BBA {} Hasselmo, M\BPBI E.%
\end{APACrefauthors}%
\unskip\
\newblock
\APACrefYearMonthDay{2002}{}{}.
\newblock
{\BBOQ}\APACrefatitle {Simulations of the Role of the Muscarinic-Activated Calcium-Sensitive Nonspecific Cation Current {INCM} in Entorhinal Neuronal Activity during Delayed Matching Tasks} {Simulations of the role of the muscarinic-activated calcium-sensitive nonspecific cation current {INCM} in entorhinal neuronal activity during delayed matching tasks}.{\BBCQ}
\newblock
\APACjournalVolNumPages{Journal of Neuroscience}{22}{3}{1081-1097}.
\PrintBackRefs{\CurrentBib}

\bibitem [\protect \citeauthoryear {%
Frans\'{e}n%
, Tahvildari%
, Egorov%
, Hasselmo%
\BCBL {}\ \BBA {} Alonso%
}{%
Frans\'{e}n%
\ \protect \BOthers {.}}{%
{\protect \APACyear {2006}}%
}]{%
FranEtal06}
\APACinsertmetastar {%
FranEtal06}%
\begin{APACrefauthors}%
Frans\'{e}n, E.%
, Tahvildari, B.%
, Egorov, A\BPBI V.%
, Hasselmo, M\BPBI E.%
\BCBL {}\ \BBA {} Alonso, A\BPBI A.%
\end{APACrefauthors}%
\unskip\
\newblock
\APACrefYearMonthDay{2006}{}{}.
\newblock
{\BBOQ}\APACrefatitle {Mechanism of graded persistent cellular activity of entorhinal cortex layer {V} neurons.} {Mechanism of graded persistent cellular activity of entorhinal cortex layer {V} neurons.}{\BBCQ}
\newblock
\APACjournalVolNumPages{Neuron}{49}{5}{735-46}.
\PrintBackRefs{\CurrentBib}

\bibitem [\protect \citeauthoryear {%
Fusi%
, Miller%
\BCBL {}\ \BBA {} Rigotti%
}{%
Fusi%
\ \protect \BOthers {.}}{%
{\protect \APACyear {2016}}%
}]{%
FusiEtal16}
\APACinsertmetastar {%
FusiEtal16}%
\begin{APACrefauthors}%
Fusi, S.%
, Miller, E\BPBI K.%
\BCBL {}\ \BBA {} Rigotti, M.%
\end{APACrefauthors}%
\unskip\
\newblock
\APACrefYearMonthDay{2016}{}{}.
\newblock
{\BBOQ}\APACrefatitle {Why neurons mix: high dimensionality for higher cognition} {Why neurons mix: high dimensionality for higher cognition}.{\BBCQ}
\newblock
\APACjournalVolNumPages{Current opinion in neurobiology}{37}{}{66--74}.
\PrintBackRefs{\CurrentBib}

\bibitem [\protect \citeauthoryear {%
Gallistel%
\ \BBA {} Gelman%
}{%
Gallistel%
\ \BBA {} Gelman%
}{%
{\protect \APACyear {1992}}%
}]{%
GallGelm92}
\APACinsertmetastar {%
GallGelm92}%
\begin{APACrefauthors}%
Gallistel, C\BPBI R.%
\BCBT {}\ \BBA {} Gelman, R.%
\end{APACrefauthors}%
\unskip\
\newblock
\APACrefYearMonthDay{1992}{}{}.
\newblock
{\BBOQ}\APACrefatitle {Preverbal and verbal counting and computation} {Preverbal and verbal counting and computation}.{\BBCQ}
\newblock
\APACjournalVolNumPages{Cognition}{44}{1}{43--74}.
\PrintBackRefs{\CurrentBib}

\bibitem [\protect \citeauthoryear {%
Grossberg%
}{%
Grossberg%
}{%
{\protect \APACyear {1969}}%
}]{%
Gros69}
\APACinsertmetastar {%
Gros69}%
\begin{APACrefauthors}%
Grossberg, S.%
\end{APACrefauthors}%
\unskip\
\newblock
\APACrefYearMonthDay{1969}{}{}.
\newblock
{\BBOQ}\APACrefatitle {On the serial learning of lists} {On the serial learning of lists}.{\BBCQ}
\newblock
\APACjournalVolNumPages{Mathematical Biosciences}{4}{1-2}{201--253}.
\PrintBackRefs{\CurrentBib}

\bibitem [\protect \citeauthoryear {%
Grossberg%
}{%
Grossberg%
}{%
{\protect \APACyear {1978}}%
}]{%
Gros78}
\APACinsertmetastar {%
Gros78}%
\begin{APACrefauthors}%
Grossberg, S.%
\end{APACrefauthors}%
\unskip\
\newblock
\APACrefYearMonthDay{1978}{}{}.
\newblock
{\BBOQ}\APACrefatitle {Behavioral contrast in short-term memory: serial binary memory models or parallel continuous memory models?} {Behavioral contrast in short-term memory: serial binary memory models or parallel continuous memory models?}{\BBCQ}
\newblock
\APACjournalVolNumPages{Journal of Mathematical Psychology}{17}{}{199-219}.
\PrintBackRefs{\CurrentBib}

\bibitem [\protect \citeauthoryear {%
Guo%
, Huson%
, Macosko%
\BCBL {}\ \BBA {} Regehr%
}{%
Guo%
\ \protect \BOthers {.}}{%
{\protect \APACyear {2021}}%
}]{%
GuoEtal21}
\APACinsertmetastar {%
GuoEtal21}%
\begin{APACrefauthors}%
Guo, C.%
, Huson, V.%
, Macosko, E\BPBI Z.%
\BCBL {}\ \BBA {} Regehr, W\BPBI G.%
\end{APACrefauthors}%
\unskip\
\newblock
\APACrefYearMonthDay{2021}{}{}.
\newblock
{\BBOQ}\APACrefatitle {Graded heterogeneity of metabotropic signaling underlies a continuum of cell-intrinsic temporal responses in unipolar brush cells} {Graded heterogeneity of metabotropic signaling underlies a continuum of cell-intrinsic temporal responses in unipolar brush cells}.{\BBCQ}
\newblock
\APACjournalVolNumPages{Nature Communications}{12}{1}{1--12}.
\PrintBackRefs{\CurrentBib}

\bibitem [\protect \citeauthoryear {%
Hacker%
}{%
Hacker%
}{%
{\protect \APACyear {1980}}%
}]{%
Hack80}
\APACinsertmetastar {%
Hack80}%
\begin{APACrefauthors}%
Hacker, M\BPBI J.%
\end{APACrefauthors}%
\unskip\
\newblock
\APACrefYearMonthDay{1980}{}{}.
\newblock
{\BBOQ}\APACrefatitle {Speed and accuracy of recency judgments for events in short-term memory.} {Speed and accuracy of recency judgments for events in short-term memory.}{\BBCQ}
\newblock
\APACjournalVolNumPages{Journal of Experimental Psychology: Human Learning and Memory}{15}{}{846-858}.
\PrintBackRefs{\CurrentBib}

\bibitem [\protect \citeauthoryear {%
Helias%
\ \BBA {} Dahmen%
}{%
Helias%
\ \BBA {} Dahmen%
}{%
{\protect \APACyear {2020}}%
}]{%
helias2020statistical}
\APACinsertmetastar {%
helias2020statistical}%
\begin{APACrefauthors}%
Helias, M.%
\BCBT {}\ \BBA {} Dahmen, D.%
\end{APACrefauthors}%
\unskip\
\newblock
\APACrefYear{2020}.
\newblock
\APACrefbtitle {Statistical field theory for neural networks} {Statistical field theory for neural networks}\ (\BVOL~970).
\newblock
\APACaddressPublisher{}{Springer}.
\PrintBackRefs{\CurrentBib}

\bibitem [\protect \citeauthoryear {%
Hintzman%
}{%
Hintzman%
}{%
{\protect \APACyear {2010}}%
}]{%
Hint10}
\APACinsertmetastar {%
Hint10}%
\begin{APACrefauthors}%
Hintzman, D\BPBI L.%
\end{APACrefauthors}%
\unskip\
\newblock
\APACrefYearMonthDay{2010}{}{}.
\newblock
{\BBOQ}\APACrefatitle {How does repetition affect memory? {Evidence} from judgments of recency.} {How does repetition affect memory? {Evidence} from judgments of recency.}{\BBCQ}
\newblock
\APACjournalVolNumPages{Memory \& Cognition}{38}{1}{102-15}.
\PrintBackRefs{\CurrentBib}

\bibitem [\protect \citeauthoryear {%
Howard%
, Esfahani%
, Le%
\BCBL {}\ \BBA {} Sederberg%
}{%
Howard%
\ \protect \BOthers {.}}{%
{\protect \APACyear {2024}}%
}]{%
HowaEtal24}
\APACinsertmetastar {%
HowaEtal24}%
\begin{APACrefauthors}%
Howard, M\BPBI W.%
, Esfahani, Z\BPBI G.%
, Le, B.%
\BCBL {}\ \BBA {} Sederberg, P\BPBI B.%
\end{APACrefauthors}%
\unskip\
\newblock
\APACrefYearMonthDay{2024}{}{}.
\newblock
{\BBOQ}\APACrefatitle {Learning temporal relationships between symbols with {Laplace Neural Manifolds}} {Learning temporal relationships between symbols with {Laplace Neural Manifolds}}.{\BBCQ}
\newblock
\APACjournalVolNumPages{Computational Brain and Behavior}{}{}{}.
\newblock
\begin{APACrefURL} \url{https://doi.org/10.1007/s42113-024-00230-8} \end{APACrefURL}
\PrintBackRefs{\CurrentBib}

\bibitem [\protect \citeauthoryear {%
Howard%
\ \BBA {} Shankar%
}{%
Howard%
\ \BBA {} Shankar%
}{%
{\protect \APACyear {2018}}%
}]{%
HowaShan18}
\APACinsertmetastar {%
HowaShan18}%
\begin{APACrefauthors}%
Howard, M\BPBI W.%
\BCBT {}\ \BBA {} Shankar, K\BPBI H.%
\end{APACrefauthors}%
\unskip\
\newblock
\APACrefYearMonthDay{2018}{}{}.
\newblock
{\BBOQ}\APACrefatitle {Neural Scaling Laws for an Uncertain World} {Neural scaling laws for an uncertain world}.{\BBCQ}
\newblock
\APACjournalVolNumPages{Psychologial Review}{125}{}{47-58}.
\newblock
\begin{APACrefDOI} \doi{10.1037/rev0000081} \end{APACrefDOI}
\PrintBackRefs{\CurrentBib}

\bibitem [\protect \citeauthoryear {%
Howard%
, Shankar%
, Aue%
\BCBL {}\ \BBA {} Criss%
}{%
Howard%
\ \protect \BOthers {.}}{%
{\protect \APACyear {2015}}%
}]{%
HowaEtal15}
\APACinsertmetastar {%
HowaEtal15}%
\begin{APACrefauthors}%
Howard, M\BPBI W.%
, Shankar, K\BPBI H.%
, Aue, W.%
\BCBL {}\ \BBA {} Criss, A\BPBI H.%
\end{APACrefauthors}%
\unskip\
\newblock
\APACrefYearMonthDay{2015}{}{}.
\newblock
{\BBOQ}\APACrefatitle {A distributed representation of internal time} {A distributed representation of internal time}.{\BBCQ}
\newblock
\APACjournalVolNumPages{Psychological Review}{122}{1}{24-53}.
\PrintBackRefs{\CurrentBib}

\bibitem [\protect \citeauthoryear {%
Husserl%
}{%
Husserl%
}{%
{\protect \APACyear {1966}}%
}]{%
Huss66}
\APACinsertmetastar {%
Huss66}%
\begin{APACrefauthors}%
Husserl, E.%
\end{APACrefauthors}%
\unskip\
\newblock
\APACrefYear{1966}.
\newblock
\APACrefbtitle {The phenomenology of internal time-consciousness} {The phenomenology of internal time-consciousness}.
\newblock
\APACaddressPublisher{Bloomington, IN}{Indiana University Press}.
\PrintBackRefs{\CurrentBib}

\bibitem [\protect \citeauthoryear {%
Jacques%
, Tiganj%
, Sarkar%
, Howard%
\BCBL {}\ \BBA {} Sederberg%
}{%
Jacques%
\ \protect \BOthers {.}}{%
{\protect \APACyear {2022}}%
}]{%
JacqEtal22}
\APACinsertmetastar {%
JacqEtal22}%
\begin{APACrefauthors}%
Jacques, B\BPBI G.%
, Tiganj, Z.%
, Sarkar, A.%
, Howard, M.%
\BCBL {}\ \BBA {} Sederberg, P.%
\end{APACrefauthors}%
\unskip\
\newblock
\APACrefYearMonthDay{2022}{}{}.
\newblock
{\BBOQ}\APACrefatitle {A deep convolutional neural network that is invariant to time rescaling} {A deep convolutional neural network that is invariant to time rescaling}.{\BBCQ}
\newblock
\BIn{} \APACrefbtitle {International Conference on Machine Learning} {International conference on machine learning}\ (\BPGS\ 9729--9738).
\PrintBackRefs{\CurrentBib}

\bibitem [\protect \citeauthoryear {%
James%
}{%
James%
}{%
{\protect \APACyear {1890}}%
}]{%
Jame90}
\APACinsertmetastar {%
Jame90}%
\begin{APACrefauthors}%
James, W.%
\end{APACrefauthors}%
\unskip\
\newblock
\APACrefYear{1890}.
\newblock
\APACrefbtitle {The principles of psychology} {The principles of psychology}.
\newblock
\APACaddressPublisher{New York}{Holt}.
\PrintBackRefs{\CurrentBib}

\bibitem [\protect \citeauthoryear {%
Jazayeri%
\ \BBA {} Ostojic%
}{%
Jazayeri%
\ \BBA {} Ostojic%
}{%
{\protect \APACyear {2021}}%
}]{%
JazaOsto21}
\APACinsertmetastar {%
JazaOsto21}%
\begin{APACrefauthors}%
Jazayeri, M.%
\BCBT {}\ \BBA {} Ostojic, S.%
\end{APACrefauthors}%
\unskip\
\newblock
\APACrefYearMonthDay{2021}{}{}.
\newblock
{\BBOQ}\APACrefatitle {Interpreting neural computations by examining intrinsic and embedding dimensionality of neural activity} {Interpreting neural computations by examining intrinsic and embedding dimensionality of neural activity}.{\BBCQ}
\newblock
\APACjournalVolNumPages{Current opinion in neurobiology}{70}{}{113--120}.
\PrintBackRefs{\CurrentBib}

\bibitem [\protect \citeauthoryear {%
Jin%
, Fujii%
\BCBL {}\ \BBA {} Graybiel%
}{%
Jin%
\ \protect \BOthers {.}}{%
{\protect \APACyear {2009}}%
}]{%
JinEtal09}
\APACinsertmetastar {%
JinEtal09}%
\begin{APACrefauthors}%
Jin, D\BPBI Z.%
, Fujii, N.%
\BCBL {}\ \BBA {} Graybiel, A\BPBI M.%
\end{APACrefauthors}%
\unskip\
\newblock
\APACrefYearMonthDay{2009}{}{}.
\newblock
{\BBOQ}\APACrefatitle {Neural representation of time in cortico-basal ganglia circuits} {Neural representation of time in cortico-basal ganglia circuits}.{\BBCQ}
\newblock
\APACjournalVolNumPages{Proceedings of the National Academy of Sciences}{106}{45}{19156--19161}.
\PrintBackRefs{\CurrentBib}

\bibitem [\protect \citeauthoryear {%
Johnston%
, Palmer%
\BCBL {}\ \BBA {} Freedman%
}{%
Johnston%
\ \protect \BOthers {.}}{%
{\protect \APACyear {2020}}%
}]{%
JohnEtal20}
\APACinsertmetastar {%
JohnEtal20}%
\begin{APACrefauthors}%
Johnston, W\BPBI J.%
, Palmer, S\BPBI E.%
\BCBL {}\ \BBA {} Freedman, D\BPBI J.%
\end{APACrefauthors}%
\unskip\
\newblock
\APACrefYearMonthDay{2020}{}{}.
\newblock
{\BBOQ}\APACrefatitle {Nonlinear mixed selectivity supports reliable neural computation} {Nonlinear mixed selectivity supports reliable neural computation}.{\BBCQ}
\newblock
\APACjournalVolNumPages{PLoS computational biology}{16}{2}{e1007544}.
\PrintBackRefs{\CurrentBib}

\bibitem [\protect \citeauthoryear {%
Kahana%
\ \BBA {} Sekuler%
}{%
Kahana%
\ \BBA {} Sekuler%
}{%
{\protect \APACyear {2002}}%
}]{%
KahaSeku02}
\APACinsertmetastar {%
KahaSeku02}%
\begin{APACrefauthors}%
Kahana, M\BPBI J.%
\BCBT {}\ \BBA {} Sekuler, R.%
\end{APACrefauthors}%
\unskip\
\newblock
\APACrefYearMonthDay{2002}{}{}.
\newblock
{\BBOQ}\APACrefatitle {Recognizing spatial patterns: a noisy exemplar approach} {Recognizing spatial patterns: a noisy exemplar approach}.{\BBCQ}
\newblock
\APACjournalVolNumPages{Vision Research}{42}{18}{2177-192}.
\PrintBackRefs{\CurrentBib}

\bibitem [\protect \citeauthoryear {%
Kim%
, Rouault%
, Druckmann%
\BCBL {}\ \BBA {} Jayaraman%
}{%
Kim%
\ \protect \BOthers {.}}{%
{\protect \APACyear {2017}}%
}]{%
KimEtal17}
\APACinsertmetastar {%
KimEtal17}%
\begin{APACrefauthors}%
Kim, S\BPBI S.%
, Rouault, H.%
, Druckmann, S.%
\BCBL {}\ \BBA {} Jayaraman, V.%
\end{APACrefauthors}%
\unskip\
\newblock
\APACrefYearMonthDay{2017}{}{}.
\newblock
{\BBOQ}\APACrefatitle {Ring attractor dynamics in the Drosophila central brain} {Ring attractor dynamics in the drosophila central brain}.{\BBCQ}
\newblock
\APACjournalVolNumPages{Science}{356}{6340}{849--853}.
\PrintBackRefs{\CurrentBib}

\bibitem [\protect \citeauthoryear {%
King%
\ \BBA {} Dehaene%
}{%
King%
\ \BBA {} Dehaene%
}{%
{\protect \APACyear {2014}}%
}]{%
KingDeha14}
\APACinsertmetastar {%
KingDeha14}%
\begin{APACrefauthors}%
King, J.%
\BCBT {}\ \BBA {} Dehaene, S.%
\end{APACrefauthors}%
\unskip\
\newblock
\APACrefYearMonthDay{2014}{}{}.
\newblock
{\BBOQ}\APACrefatitle {Characterizing the dynamics of mental representations: the temporal generalization method} {Characterizing the dynamics of mental representations: the temporal generalization method}.{\BBCQ}
\newblock
\APACjournalVolNumPages{Trends in cognitive sciences}{18}{4}{203--210}.
\PrintBackRefs{\CurrentBib}

\bibitem [\protect \citeauthoryear {%
Kirschhock%
\ \BBA {} Nieder%
}{%
Kirschhock%
\ \BBA {} Nieder%
}{%
{\protect \APACyear {2023}}%
}]{%
KirsNied23}
\APACinsertmetastar {%
KirsNied23}%
\begin{APACrefauthors}%
Kirschhock, M\BPBI E.%
\BCBT {}\ \BBA {} Nieder, A.%
\end{APACrefauthors}%
\unskip\
\newblock
\APACrefYearMonthDay{2023}{}{}.
\newblock
{\BBOQ}\APACrefatitle {Numerical representation for action in crows obeys the Weber-Fechner law} {Numerical representation for action in crows obeys the weber-fechner law}.{\BBCQ}
\newblock
\APACjournalVolNumPages{Psychological Science}{34}{12}{1322--1335}.
\PrintBackRefs{\CurrentBib}

\bibitem [\protect \citeauthoryear {%
Kohli%
, Cloninger%
\BCBL {}\ \BBA {} Mishne%
}{%
Kohli%
\ \protect \BOthers {.}}{%
{\protect \APACyear {2021}}%
}]{%
KohlEtal21}
\APACinsertmetastar {%
KohlEtal21}%
\begin{APACrefauthors}%
Kohli, D.%
, Cloninger, A.%
\BCBL {}\ \BBA {} Mishne, G.%
\end{APACrefauthors}%
\unskip\
\newblock
\APACrefYearMonthDay{2021}{}{}.
\newblock
{\BBOQ}\APACrefatitle {{LDLE: L}ow distortion local eigenmaps} {{LDLE: L}ow distortion local eigenmaps}.{\BBCQ}
\newblock
\APACjournalVolNumPages{Journal of machine learning research}{22}{282}{1--64}.
\PrintBackRefs{\CurrentBib}

\bibitem [\protect \citeauthoryear {%
Krishnamurthy%
, Can%
\BCBL {}\ \BBA {} Schwab%
}{%
Krishnamurthy%
\ \protect \BOthers {.}}{%
{\protect \APACyear {2022}}%
}]{%
KrisEtal22}
\APACinsertmetastar {%
KrisEtal22}%
\begin{APACrefauthors}%
Krishnamurthy, K.%
, Can, T.%
\BCBL {}\ \BBA {} Schwab, D\BPBI J.%
\end{APACrefauthors}%
\unskip\
\newblock
\APACrefYearMonthDay{2022}{}{}.
\newblock
{\BBOQ}\APACrefatitle {Theory of gating in recurrent neural networks} {Theory of gating in recurrent neural networks}.{\BBCQ}
\newblock
\APACjournalVolNumPages{Physical Review X}{12}{1}{011011}.
\PrintBackRefs{\CurrentBib}

\bibitem [\protect \citeauthoryear {%
Langdon%
, Genkin%
\BCBL {}\ \BBA {} Engel%
}{%
Langdon%
\ \protect \BOthers {.}}{%
{\protect \APACyear {2023}}%
}]{%
LangEtal23}
\APACinsertmetastar {%
LangEtal23}%
\begin{APACrefauthors}%
Langdon, C.%
, Genkin, M.%
\BCBL {}\ \BBA {} Engel, T\BPBI A.%
\end{APACrefauthors}%
\unskip\
\newblock
\APACrefYearMonthDay{2023}{}{}.
\newblock
{\BBOQ}\APACrefatitle {A unifying perspective on neural manifolds and circuits for cognition} {A unifying perspective on neural manifolds and circuits for cognition}.{\BBCQ}
\newblock
\APACjournalVolNumPages{Nature Reviews Neuroscience}{}{}{1--15}.
\PrintBackRefs{\CurrentBib}

\bibitem [\protect \citeauthoryear {%
Lebedev%
\ \protect \BOthers {.}}{%
Lebedev%
\ \protect \BOthers {.}}{%
{\protect \APACyear {2019}}%
}]{%
LebeEtal19}
\APACinsertmetastar {%
LebeEtal19}%
\begin{APACrefauthors}%
Lebedev, M\BPBI A.%
, Ossadtchi, A.%
, Mill, N\BPBI A.%
, Urp{\'\i}, N\BPBI A.%
, Cervera, M\BPBI R.%
\BCBL {}\ \BBA {} Nicolelis, M\BPBI A.%
\end{APACrefauthors}%
\unskip\
\newblock
\APACrefYearMonthDay{2019}{}{}.
\newblock
{\BBOQ}\APACrefatitle {What, if anything, is the true neurophysiological significance of rotational dynamics?} {What, if anything, is the true neurophysiological significance of rotational dynamics?}{\BBCQ}
\newblock
\APACjournalVolNumPages{BioRxiv}{}{}{597419}.
\PrintBackRefs{\CurrentBib}

\bibitem [\protect \citeauthoryear {%
Liu%
\ \BBA {} Howard%
}{%
Liu%
\ \BBA {} Howard%
}{%
{\protect \APACyear {2020}}%
}]{%
LiuHowa20}
\APACinsertmetastar {%
LiuHowa20}%
\begin{APACrefauthors}%
Liu, Y.%
\BCBT {}\ \BBA {} Howard, M\BPBI W.%
\end{APACrefauthors}%
\unskip\
\newblock
\APACrefYearMonthDay{2020}{}{}.
\newblock
{\BBOQ}\APACrefatitle {Generation of scale-invariant sequential activity in linear recurrent networks} {Generation of scale-invariant sequential activity in linear recurrent networks}.{\BBCQ}
\newblock
\APACjournalVolNumPages{Neural Computation}{32}{7}{1379--1407}.
\PrintBackRefs{\CurrentBib}

\bibitem [\protect \citeauthoryear {%
Loewenstein%
\ \BBA {} Sompolinsky%
}{%
Loewenstein%
\ \BBA {} Sompolinsky%
}{%
{\protect \APACyear {2003}}%
}]{%
LoewSomp03}
\APACinsertmetastar {%
LoewSomp03}%
\begin{APACrefauthors}%
Loewenstein, Y.%
\BCBT {}\ \BBA {} Sompolinsky, H.%
\end{APACrefauthors}%
\unskip\
\newblock
\APACrefYearMonthDay{2003}{}{}.
\newblock
{\BBOQ}\APACrefatitle {Temporal integration by calcium dynamics in a model neuron.} {Temporal integration by calcium dynamics in a model neuron.}{\BBCQ}
\newblock
\APACjournalVolNumPages{Nature Neuroscience}{6}{9}{961-7}.
\PrintBackRefs{\CurrentBib}

\bibitem [\protect \citeauthoryear {%
Luce%
\ \BBA {} Suppes%
}{%
Luce%
\ \BBA {} Suppes%
}{%
{\protect \APACyear {2002}}%
}]{%
LuceSupp02}
\APACinsertmetastar {%
LuceSupp02}%
\begin{APACrefauthors}%
Luce, R\BPBI D.%
\BCBT {}\ \BBA {} Suppes, P.%
\end{APACrefauthors}%
\unskip\
\newblock
\APACrefYearMonthDay{2002}{}{}.
\newblock
{\BBOQ}\APACrefatitle {Representational measurement theory} {Representational measurement theory}.{\BBCQ}
\newblock
\BIn{} J.~Wixted\ \BBA {} H.~Pashler\ (\BEDS), \APACrefbtitle {Stevens Handbook of Experimental Psychology, 3rd Edition} {Stevens handbook of experimental psychology, 3rd edition}\ (\BVOL\ 4: Methodology in Experimental Psychology, \BPG~1-41).
\newblock
\APACaddressPublisher{}{Wiley Online Library}.
\PrintBackRefs{\CurrentBib}

\bibitem [\protect \citeauthoryear {%
Luck%
\ \BBA {} Vogel%
}{%
Luck%
\ \BBA {} Vogel%
}{%
{\protect \APACyear {1997}}%
}]{%
LuckVoge97}
\APACinsertmetastar {%
LuckVoge97}%
\begin{APACrefauthors}%
Luck, S\BPBI J.%
\BCBT {}\ \BBA {} Vogel, E\BPBI K.%
\end{APACrefauthors}%
\unskip\
\newblock
\APACrefYearMonthDay{1997}{}{}.
\newblock
{\BBOQ}\APACrefatitle {The capacity of visual working memory for features and conjunctions.} {The capacity of visual working memory for features and conjunctions.}{\BBCQ}
\newblock
\APACjournalVolNumPages{Nature}{390}{6657}{279-81}.
\PrintBackRefs{\CurrentBib}

\bibitem [\protect \citeauthoryear {%
Lundqvist%
, Herman%
\BCBL {}\ \BBA {} Miller%
}{%
Lundqvist%
\ \protect \BOthers {.}}{%
{\protect \APACyear {2018}}%
}]{%
LundEtal18a}
\APACinsertmetastar {%
LundEtal18a}%
\begin{APACrefauthors}%
Lundqvist, M.%
, Herman, P.%
\BCBL {}\ \BBA {} Miller, E\BPBI K.%
\end{APACrefauthors}%
\unskip\
\newblock
\APACrefYearMonthDay{2018}{}{}.
\newblock
{\BBOQ}\APACrefatitle {Working Memory: Delay Activity, Yes! Persistent Activity? Maybe Not} {Working memory: Delay activity, yes! persistent activity? maybe not}.{\BBCQ}
\newblock
\APACjournalVolNumPages{Journal of Neuroscience}{38}{32}{7013-7019}.
\newblock
\begin{APACrefDOI} \doi{10.1523/JNEUROSCI.2485-17.2018} \end{APACrefDOI}
\PrintBackRefs{\CurrentBib}

\bibitem [\protect \citeauthoryear {%
Ma%
, Husain%
\BCBL {}\ \BBA {} Bays%
}{%
Ma%
\ \protect \BOthers {.}}{%
{\protect \APACyear {2014}}%
}]{%
MaEtal14}
\APACinsertmetastar {%
MaEtal14}%
\begin{APACrefauthors}%
Ma, W\BPBI J.%
, Husain, M.%
\BCBL {}\ \BBA {} Bays, P\BPBI M.%
\end{APACrefauthors}%
\unskip\
\newblock
\APACrefYearMonthDay{2014}{}{}.
\newblock
{\BBOQ}\APACrefatitle {Changing concepts of working memory} {Changing concepts of working memory}.{\BBCQ}
\newblock
\APACjournalVolNumPages{Nature neuroscience}{17}{3}{347--356}.
\PrintBackRefs{\CurrentBib}

\bibitem [\protect \citeauthoryear {%
Maass%
, Natschl{\"a}ger%
\BCBL {}\ \BBA {} Markram%
}{%
Maass%
\ \protect \BOthers {.}}{%
{\protect \APACyear {2002}}%
}]{%
MaasEtal02}
\APACinsertmetastar {%
MaasEtal02}%
\begin{APACrefauthors}%
Maass, W.%
, Natschl{\"a}ger, T.%
\BCBL {}\ \BBA {} Markram, H.%
\end{APACrefauthors}%
\unskip\
\newblock
\APACrefYearMonthDay{2002}{}{}.
\newblock
{\BBOQ}\APACrefatitle {Real-time computing without stable states: a new framework for neural computation based on perturbations} {Real-time computing without stable states: a new framework for neural computation based on perturbations}.{\BBCQ}
\newblock
\APACjournalVolNumPages{Neural Computation}{14}{11}{2531-60}.
\newblock
\begin{APACrefDOI} \doi{10.1162/089976602760407955} \end{APACrefDOI}
\PrintBackRefs{\CurrentBib}

\bibitem [\protect \citeauthoryear {%
Mac{D}onald%
, Lepage%
, Eden%
\BCBL {}\ \BBA {} Eichenbaum%
}{%
Mac{D}onald%
\ \protect \BOthers {.}}{%
{\protect \APACyear {2011}}%
}]{%
MacDEtal11}
\APACinsertmetastar {%
MacDEtal11}%
\begin{APACrefauthors}%
Mac{D}onald, C\BPBI J.%
, Lepage, K\BPBI Q.%
, Eden, U\BPBI T.%
\BCBL {}\ \BBA {} Eichenbaum, H.%
\end{APACrefauthors}%
\unskip\
\newblock
\APACrefYearMonthDay{2011}{}{}.
\newblock
{\BBOQ}\APACrefatitle {Hippocampal ``Time Cells'' Bridge the Gap in Memory for Discontiguous Events} {Hippocampal ``time cells'' bridge the gap in memory for discontiguous events}.{\BBCQ}
\newblock
\APACjournalVolNumPages{Neuron}{71}{4}{737-749}.
\PrintBackRefs{\CurrentBib}

\bibitem [\protect \citeauthoryear {%
Machens%
, Romo%
\BCBL {}\ \BBA {} Brody%
}{%
Machens%
\ \protect \BOthers {.}}{%
{\protect \APACyear {2010}}%
}]{%
MachEtal10}
\APACinsertmetastar {%
MachEtal10}%
\begin{APACrefauthors}%
Machens, C\BPBI K.%
, Romo, R.%
\BCBL {}\ \BBA {} Brody, C\BPBI D.%
\end{APACrefauthors}%
\unskip\
\newblock
\APACrefYearMonthDay{2010}{}{}.
\newblock
{\BBOQ}\APACrefatitle {Functional, but not anatomical, separation of what and when in prefrontal cortex} {Functional, but not anatomical, separation of what and when in prefrontal cortex}.{\BBCQ}
\newblock
\APACjournalVolNumPages{Journal of Neuroscience}{30}{1}{350--360}.
\PrintBackRefs{\CurrentBib}

\bibitem [\protect \citeauthoryear {%
Manley%
\ \protect \BOthers {.}}{%
Manley%
\ \protect \BOthers {.}}{%
{\protect \APACyear {2024}}%
}]{%
ManlEtal24}
\APACinsertmetastar {%
ManlEtal24}%
\begin{APACrefauthors}%
Manley, J.%
, Lu, S.%
, Barber, K.%
, Demas, J.%
, Kim, H.%
, Meyer, D.%
\BDBL {}Vaziri, A.%
\end{APACrefauthors}%
\unskip\
\newblock
\APACrefYearMonthDay{2024}{}{}.
\newblock
{\BBOQ}\APACrefatitle {Simultaneous, cortex-wide dynamics of up to 1 million neurons reveal unbounded scaling of dimensionality with neuron number} {Simultaneous, cortex-wide dynamics of up to 1 million neurons reveal unbounded scaling of dimensionality with neuron number}.{\BBCQ}
\newblock
\APACjournalVolNumPages{Neuron}{112}{10}{1694--1709}.
\PrintBackRefs{\CurrentBib}

\bibitem [\protect \citeauthoryear {%
Mante%
, Sussillo%
, Shenoy%
\BCBL {}\ \BBA {} Newsome%
}{%
Mante%
\ \protect \BOthers {.}}{%
{\protect \APACyear {2013}}%
}]{%
MantEtal13}
\APACinsertmetastar {%
MantEtal13}%
\begin{APACrefauthors}%
Mante, V.%
, Sussillo, D.%
, Shenoy, K\BPBI V.%
\BCBL {}\ \BBA {} Newsome, W\BPBI T.%
\end{APACrefauthors}%
\unskip\
\newblock
\APACrefYearMonthDay{2013}{}{}.
\newblock
{\BBOQ}\APACrefatitle {Context-dependent computation by recurrent dynamics in prefrontal cortex} {Context-dependent computation by recurrent dynamics in prefrontal cortex}.{\BBCQ}
\newblock
\APACjournalVolNumPages{Nature}{503}{7474}{78--84}.
\PrintBackRefs{\CurrentBib}

\bibitem [\protect \citeauthoryear {%
McInnes%
, Healy%
\BCBL {}\ \BBA {} Melville%
}{%
McInnes%
\ \protect \BOthers {.}}{%
{\protect \APACyear {2018}}%
}]{%
McInEtal18}
\APACinsertmetastar {%
McInEtal18}%
\begin{APACrefauthors}%
McInnes, L.%
, Healy, J.%
\BCBL {}\ \BBA {} Melville, J.%
\end{APACrefauthors}%
\unskip\
\newblock
\APACrefYearMonthDay{2018}{}{}.
\newblock
{\BBOQ}\APACrefatitle {Umap: Uniform manifold approximation and projection for dimension reduction} {Umap: Uniform manifold approximation and projection for dimension reduction}.{\BBCQ}
\newblock
\APACjournalVolNumPages{arXiv preprint arXiv:1802.03426}{}{}{}.
\PrintBackRefs{\CurrentBib}

\bibitem [\protect \citeauthoryear {%
Murdock%
}{%
Murdock%
}{%
{\protect \APACyear {1960}}%
}]{%
Murd60}
\APACinsertmetastar {%
Murd60}%
\begin{APACrefauthors}%
Murdock, B\BPBI B.%
\end{APACrefauthors}%
\unskip\
\newblock
\APACrefYearMonthDay{1960}{}{}.
\newblock
{\BBOQ}\APACrefatitle {The distinctiveness of stimuli} {The distinctiveness of stimuli}.{\BBCQ}
\newblock
\APACjournalVolNumPages{Psychological Review}{67}{}{16-31}.
\PrintBackRefs{\CurrentBib}

\bibitem [\protect \citeauthoryear {%
Murray%
\ \protect \BOthers {.}}{%
Murray%
\ \protect \BOthers {.}}{%
{\protect \APACyear {2017}}%
}]{%
MurrEtal17}
\APACinsertmetastar {%
MurrEtal17}%
\begin{APACrefauthors}%
Murray, J\BPBI D.%
, Bernacchia, A.%
, Roy, N\BPBI A.%
, Constantinidis, C.%
, Romo, R.%
\BCBL {}\ \BBA {} Wang, X\BHBI J.%
\end{APACrefauthors}%
\unskip\
\newblock
\APACrefYearMonthDay{2017}{dec}{}.
\newblock
{\BBOQ}\APACrefatitle {Stable population coding for working memory coexists with heterogeneous neural dynamics in prefrontal cortex} {Stable population coding for working memory coexists with heterogeneous neural dynamics in prefrontal cortex}.{\BBCQ}
\newblock
\APACjournalVolNumPages{Proceedings of the National Academy of Sciences}{114}{2}{394--399}.
\newblock
\begin{APACrefURL} \url{https://doi.org/10.1073/pnas.1619449114} \end{APACrefURL}
\newblock
\begin{APACrefDOI} \doi{10.1073/pnas.1619449114} \end{APACrefDOI}
\PrintBackRefs{\CurrentBib}

\bibitem [\protect \citeauthoryear {%
Nieder%
\ \BBA {} Dehaene%
}{%
Nieder%
\ \BBA {} Dehaene%
}{%
{\protect \APACyear {2009}}%
}]{%
NiedDeha09}
\APACinsertmetastar {%
NiedDeha09}%
\begin{APACrefauthors}%
Nieder, A.%
\BCBT {}\ \BBA {} Dehaene, S.%
\end{APACrefauthors}%
\unskip\
\newblock
\APACrefYearMonthDay{2009}{}{}.
\newblock
{\BBOQ}\APACrefatitle {Representation of number in the brain} {Representation of number in the brain}.{\BBCQ}
\newblock
\APACjournalVolNumPages{Annual Review of Neuroscience}{32}{}{185-208}.
\newblock
\begin{APACrefDOI} \doi{10.1146/annurev.neuro.051508.135550} \end{APACrefDOI}
\PrintBackRefs{\CurrentBib}

\bibitem [\protect \citeauthoryear {%
Nieder%
\ \BBA {} Miller%
}{%
Nieder%
\ \BBA {} Miller%
}{%
{\protect \APACyear {2003}}%
}]{%
NiedMill03}
\APACinsertmetastar {%
NiedMill03}%
\begin{APACrefauthors}%
Nieder, A.%
\BCBT {}\ \BBA {} Miller, E\BPBI K.%
\end{APACrefauthors}%
\unskip\
\newblock
\APACrefYearMonthDay{2003}{}{}.
\newblock
{\BBOQ}\APACrefatitle {Coding of cognitive magnitude: compressed scaling of numerical information in the primate prefrontal cortex} {Coding of cognitive magnitude: compressed scaling of numerical information in the primate prefrontal cortex}.{\BBCQ}
\newblock
\APACjournalVolNumPages{Neuron}{37}{1}{149-57}.
\PrintBackRefs{\CurrentBib}

\bibitem [\protect \citeauthoryear {%
Nieh%
\ \protect \BOthers {.}}{%
Nieh%
\ \protect \BOthers {.}}{%
{\protect \APACyear {2021}}%
}]{%
NiehEtal21}
\APACinsertmetastar {%
NiehEtal21}%
\begin{APACrefauthors}%
Nieh, E\BPBI H.%
, Schottdorf, M.%
, Freeman, N\BPBI W.%
, Low, R\BPBI J.%
, Lewallen, S.%
, Koay, S\BPBI A.%
\BDBL {}Tank, D\BPBI W.%
\end{APACrefauthors}%
\unskip\
\newblock
\APACrefYearMonthDay{2021}{}{}.
\newblock
{\BBOQ}\APACrefatitle {Geometry of abstract learned knowledge in the hippocampus} {Geometry of abstract learned knowledge in the hippocampus}.{\BBCQ}
\newblock
\APACjournalVolNumPages{Nature}{595}{7865}{80--84}.
\PrintBackRefs{\CurrentBib}

\bibitem [\protect \citeauthoryear {%
Parker%
\ \protect \BOthers {.}}{%
Parker%
\ \protect \BOthers {.}}{%
{\protect \APACyear {2022}}%
}]{%
ParkEtal22}
\APACinsertmetastar {%
ParkEtal22}%
\begin{APACrefauthors}%
Parker, N\BPBI F.%
, Baidya, A.%
, Cox, J.%
, Haetzel, L\BPBI M.%
, Zhukovskaya, A.%
, Murugan, M.%
\BDBL {}Witten, I\BPBI B.%
\end{APACrefauthors}%
\unskip\
\newblock
\APACrefYearMonthDay{2022}{}{}.
\newblock
{\BBOQ}\APACrefatitle {Choice-selective sequences dominate in cortical relative to thalamic inputs to NAc to support reinforcement learning} {Choice-selective sequences dominate in cortical relative to thalamic inputs to nac to support reinforcement learning}.{\BBCQ}
\newblock
\APACjournalVolNumPages{Cell Reports}{39}{7}{110756}.
\PrintBackRefs{\CurrentBib}

\bibitem [\protect \citeauthoryear {%
Pastalkova%
, Itskov%
, Amarasingham%
\BCBL {}\ \BBA {} Buzsaki%
}{%
Pastalkova%
\ \protect \BOthers {.}}{%
{\protect \APACyear {2008}}%
}]{%
PastEtal08}
\APACinsertmetastar {%
PastEtal08}%
\begin{APACrefauthors}%
Pastalkova, E.%
, Itskov, V.%
, Amarasingham, A.%
\BCBL {}\ \BBA {} Buzsaki, G.%
\end{APACrefauthors}%
\unskip\
\newblock
\APACrefYearMonthDay{2008}{}{}.
\newblock
{\BBOQ}\APACrefatitle {Internally generated cell assembly sequences in the rat hippocampus.} {Internally generated cell assembly sequences in the rat hippocampus.}{\BBCQ}
\newblock
\APACjournalVolNumPages{Science}{321}{5894}{1322-7}.
\PrintBackRefs{\CurrentBib}

\bibitem [\protect \citeauthoryear {%
Piantadosi%
}{%
Piantadosi%
}{%
{\protect \APACyear {2016}}%
}]{%
Pian16}
\APACinsertmetastar {%
Pian16}%
\begin{APACrefauthors}%
Piantadosi, S\BPBI T.%
\end{APACrefauthors}%
\unskip\
\newblock
\APACrefYearMonthDay{2016}{}{}.
\newblock
{\BBOQ}\APACrefatitle {A rational analysis of the approximate number system} {A rational analysis of the approximate number system}.{\BBCQ}
\newblock
\APACjournalVolNumPages{Psychonomic Bulletin \& Review}{}{}{1--10}.
\PrintBackRefs{\CurrentBib}

\bibitem [\protect \citeauthoryear {%
Posani%
, Wang%
, Muscinelli%
, Paninski%
\BCBL {}\ \BBA {} Fusi%
}{%
Posani%
\ \protect \BOthers {.}}{%
{\protect \APACyear {2024}}%
}]{%
PosaEtal24}
\APACinsertmetastar {%
PosaEtal24}%
\begin{APACrefauthors}%
Posani, L.%
, Wang, S.%
, Muscinelli, S.%
, Paninski, L.%
\BCBL {}\ \BBA {} Fusi, S.%
\end{APACrefauthors}%
\unskip\
\newblock
\APACrefYearMonthDay{2024}{}{}.
\newblock
{\BBOQ}\APACrefatitle {Rarely categorical, always high-dimensional: how the neural code changes along the cortical hierarchy} {Rarely categorical, always high-dimensional: how the neural code changes along the cortical hierarchy}.{\BBCQ}
\newblock
\APACjournalVolNumPages{bioRxiv}{}{}{2024--11}.
\PrintBackRefs{\CurrentBib}

\bibitem [\protect \citeauthoryear {%
Pouget%
\ \BBA {} Sejnowski%
}{%
Pouget%
\ \BBA {} Sejnowski%
}{%
{\protect \APACyear {1997}}%
}]{%
PougSejn97}
\APACinsertmetastar {%
PougSejn97}%
\begin{APACrefauthors}%
Pouget, A.%
\BCBT {}\ \BBA {} Sejnowski, T\BPBI J.%
\end{APACrefauthors}%
\unskip\
\newblock
\APACrefYearMonthDay{1997}{}{}.
\newblock
{\BBOQ}\APACrefatitle {Spatial transformations in the parietal cortex using basis functions} {Spatial transformations in the parietal cortex using basis functions}.{\BBCQ}
\newblock
\APACjournalVolNumPages{Journal of cognitive neuroscience}{9}{2}{222--237}.
\PrintBackRefs{\CurrentBib}

\bibitem [\protect \citeauthoryear {%
Redish%
, Elga%
\BCBL {}\ \BBA {} Touretzky%
}{%
Redish%
\ \protect \BOthers {.}}{%
{\protect \APACyear {1996}}%
}]{%
RediEtal96}
\APACinsertmetastar {%
RediEtal96}%
\begin{APACrefauthors}%
Redish, A\BPBI D.%
, Elga, A\BPBI N.%
\BCBL {}\ \BBA {} Touretzky, D\BPBI S.%
\end{APACrefauthors}%
\unskip\
\newblock
\APACrefYearMonthDay{1996}{}{}.
\newblock
{\BBOQ}\APACrefatitle {A coupled attractor model of the rodent head direction system} {A coupled attractor model of the rodent head direction system}.{\BBCQ}
\newblock
\APACjournalVolNumPages{Network: Computation in Neural Systems}{7}{4}{671-685}.
\PrintBackRefs{\CurrentBib}

\bibitem [\protect \citeauthoryear {%
Rigotti%
\ \protect \BOthers {.}}{%
Rigotti%
\ \protect \BOthers {.}}{%
{\protect \APACyear {2013}}%
}]{%
RigoEtal13}
\APACinsertmetastar {%
RigoEtal13}%
\begin{APACrefauthors}%
Rigotti, M.%
, Barak, O.%
, Warden, M\BPBI R.%
, Wang, X\BHBI J.%
, Daw, N\BPBI D.%
, Miller, E\BPBI K.%
\BCBL {}\ \BBA {} Fusi, S.%
\end{APACrefauthors}%
\unskip\
\newblock
\APACrefYearMonthDay{2013}{}{}.
\newblock
{\BBOQ}\APACrefatitle {The importance of mixed selectivity in complex cognitive tasks} {The importance of mixed selectivity in complex cognitive tasks}.{\BBCQ}
\newblock
\APACjournalVolNumPages{Nature}{497}{7451}{585-90}.
\newblock
\begin{APACrefDOI} \doi{10.1038/nature12160} \end{APACrefDOI}
\PrintBackRefs{\CurrentBib}

\bibitem [\protect \citeauthoryear {%
Roweis%
\ \BBA {} Saul%
}{%
Roweis%
\ \BBA {} Saul%
}{%
{\protect \APACyear {2000}}%
}]{%
RoweSaul00}
\APACinsertmetastar {%
RoweSaul00}%
\begin{APACrefauthors}%
Roweis, S\BPBI T.%
\BCBT {}\ \BBA {} Saul, L\BPBI K.%
\end{APACrefauthors}%
\unskip\
\newblock
\APACrefYearMonthDay{2000}{}{}.
\newblock
{\BBOQ}\APACrefatitle {Nonlinear dimensionality reduction by locally linear embedding} {Nonlinear dimensionality reduction by locally linear embedding}.{\BBCQ}
\newblock
\APACjournalVolNumPages{Science}{290}{5500}{2323-6}.
\newblock
\begin{APACrefDOI} \doi{10.1126/science.290.5500.2323} \end{APACrefDOI}
\PrintBackRefs{\CurrentBib}

\bibitem [\protect \citeauthoryear {%
Saber~Marouf%
\ \protect \BOthers {.}}{%
Saber~Marouf%
\ \protect \BOthers {.}}{%
{\protect \APACyear {2025}}%
}]{%
MaroEtal25}
\APACinsertmetastar {%
MaroEtal25}%
\begin{APACrefauthors}%
Saber~Marouf, B.%
, Reboreda, A.%
, Theissen, F.%
, Kaushik, R.%
, Sauvage, M.%
, Dityatev, A.%
\BCBL {}\ \BBA {} Yoshida, M.%
\end{APACrefauthors}%
\unskip\
\newblock
\APACrefYearMonthDay{2025}{}{}.
\newblock
{\BBOQ}\APACrefatitle {Single neurons act as a memory buffer for space} {Single neurons act as a memory buffer for space}.{\BBCQ}
\newblock
\APACjournalVolNumPages{bioRxiv}{}{}{2025--06}.
\PrintBackRefs{\CurrentBib}

\bibitem [\protect \citeauthoryear {%
S{\'a}godi%
, Mart{\'\i}n-S{\'a}nchez%
, Sok{\'o}{\l}%
\BCBL {}\ \BBA {} Park%
}{%
S{\'a}godi%
\ \protect \BOthers {.}}{%
{\protect \APACyear {2024}}%
}]{%
SagoEtal24}
\APACinsertmetastar {%
SagoEtal24}%
\begin{APACrefauthors}%
S{\'a}godi, {\'A}.%
, Mart{\'\i}n-S{\'a}nchez, G.%
, Sok{\'o}{\l}, P.%
\BCBL {}\ \BBA {} Park, I\BPBI M.%
\end{APACrefauthors}%
\unskip\
\newblock
\APACrefYearMonthDay{2024}{}{}.
\newblock
{\BBOQ}\APACrefatitle {Back to the Continuous Attractor} {Back to the continuous attractor}.{\BBCQ}
\newblock
\APACjournalVolNumPages{ArXiv}{}{}{}.
\PrintBackRefs{\CurrentBib}

\bibitem [\protect \citeauthoryear {%
Salinas%
\ \BBA {} Abbott%
}{%
Salinas%
\ \BBA {} Abbott%
}{%
{\protect \APACyear {2001}}%
}]{%
SaliAbbo01}
\APACinsertmetastar {%
SaliAbbo01}%
\begin{APACrefauthors}%
Salinas, E.%
\BCBT {}\ \BBA {} Abbott, L.%
\end{APACrefauthors}%
\unskip\
\newblock
\APACrefYearMonthDay{2001}{}{}.
\newblock
{\BBOQ}\APACrefatitle {Coordinate transformations in the visual system: how to generate gain fields and what to compute with them} {Coordinate transformations in the visual system: how to generate gain fields and what to compute with them}.{\BBCQ}
\newblock
\APACjournalVolNumPages{Progress in brain research}{130}{}{175--190}.
\PrintBackRefs{\CurrentBib}

\bibitem [\protect \citeauthoryear {%
Schurgin%
, Wixted%
\BCBL {}\ \BBA {} Brady%
}{%
Schurgin%
\ \protect \BOthers {.}}{%
{\protect \APACyear {2020}}%
}]{%
SchuEtal20}
\APACinsertmetastar {%
SchuEtal20}%
\begin{APACrefauthors}%
Schurgin, M\BPBI W.%
, Wixted, J\BPBI T.%
\BCBL {}\ \BBA {} Brady, T\BPBI F.%
\end{APACrefauthors}%
\unskip\
\newblock
\APACrefYearMonthDay{2020}{}{}.
\newblock
{\BBOQ}\APACrefatitle {Psychophysical scaling reveals a unified theory of visual memory strength} {Psychophysical scaling reveals a unified theory of visual memory strength}.{\BBCQ}
\newblock
\APACjournalVolNumPages{Nature Human Behaviour}{4}{11}{1156--1172}.
\PrintBackRefs{\CurrentBib}

\bibitem [\protect \citeauthoryear {%
Shankar%
\ \BBA {} Howard%
}{%
Shankar%
\ \BBA {} Howard%
}{%
{\protect \APACyear {2012}}%
}]{%
ShanHowa12}
\APACinsertmetastar {%
ShanHowa12}%
\begin{APACrefauthors}%
Shankar, K\BPBI H.%
\BCBT {}\ \BBA {} Howard, M\BPBI W.%
\end{APACrefauthors}%
\unskip\
\newblock
\APACrefYearMonthDay{2012}{}{}.
\newblock
{\BBOQ}\APACrefatitle {A scale-invariant internal representation of time} {A scale-invariant internal representation of time}.{\BBCQ}
\newblock
\APACjournalVolNumPages{Neural Computation}{24}{1}{134-193}.
\PrintBackRefs{\CurrentBib}

\bibitem [\protect \citeauthoryear {%
Shankar%
\ \BBA {} Howard%
}{%
Shankar%
\ \BBA {} Howard%
}{%
{\protect \APACyear {2013}}%
}]{%
ShanHowa13}
\APACinsertmetastar {%
ShanHowa13}%
\begin{APACrefauthors}%
Shankar, K\BPBI H.%
\BCBT {}\ \BBA {} Howard, M\BPBI W.%
\end{APACrefauthors}%
\unskip\
\newblock
\APACrefYearMonthDay{2013}{}{}.
\newblock
{\BBOQ}\APACrefatitle {Optimally fuzzy temporal memory} {Optimally fuzzy temporal memory}.{\BBCQ}
\newblock
\APACjournalVolNumPages{Journal of Machine Learning Research}{14}{}{3753-3780}.
\PrintBackRefs{\CurrentBib}

\bibitem [\protect \citeauthoryear {%
Shin%
, Zou%
\BCBL {}\ \BBA {} Ma%
}{%
Shin%
\ \protect \BOthers {.}}{%
{\protect \APACyear {2017}}%
}]{%
ShinEtal17}
\APACinsertmetastar {%
ShinEtal17}%
\begin{APACrefauthors}%
Shin, H.%
, Zou, Q.%
\BCBL {}\ \BBA {} Ma, W\BPBI J.%
\end{APACrefauthors}%
\unskip\
\newblock
\APACrefYearMonthDay{2017}{}{}.
\newblock
{\BBOQ}\APACrefatitle {The effects of delay duration on visual working memory for orientation} {The effects of delay duration on visual working memory for orientation}.{\BBCQ}
\newblock
\APACjournalVolNumPages{Journal of vision}{17}{14}{10--10}.
\PrintBackRefs{\CurrentBib}

\bibitem [\protect \citeauthoryear {%
Shinn%
}{%
Shinn%
}{%
{\protect \APACyear {2023}}%
}]{%
Shin23}
\APACinsertmetastar {%
Shin23}%
\begin{APACrefauthors}%
Shinn, M.%
\end{APACrefauthors}%
\unskip\
\newblock
\APACrefYearMonthDay{2023}{}{}.
\newblock
{\BBOQ}\APACrefatitle {Phantom oscillations in principal component analysis} {Phantom oscillations in principal component analysis}.{\BBCQ}
\newblock
\APACjournalVolNumPages{Proceedings of the National Academy of Sciences}{120}{48}{e2311420120}.
\PrintBackRefs{\CurrentBib}

\bibitem [\protect \citeauthoryear {%
Stevens%
}{%
Stevens%
}{%
{\protect \APACyear {1957}}%
}]{%
Stev57}
\APACinsertmetastar {%
Stev57}%
\begin{APACrefauthors}%
Stevens, S\BPBI S.%
\end{APACrefauthors}%
\unskip\
\newblock
\APACrefYearMonthDay{1957}{}{}.
\newblock
{\BBOQ}\APACrefatitle {On the psychophysical law} {On the psychophysical law}.{\BBCQ}
\newblock
\APACjournalVolNumPages{Psychological Review}{64}{3}{153-81}.
\PrintBackRefs{\CurrentBib}

\bibitem [\protect \citeauthoryear {%
Subramanian%
\ \BBA {} Smith%
}{%
Subramanian%
\ \BBA {} Smith%
}{%
{\protect \APACyear {2024}}%
}]{%
SubrSmit24}
\APACinsertmetastar {%
SubrSmit24}%
\begin{APACrefauthors}%
Subramanian, D\BPBI L.%
\BCBT {}\ \BBA {} Smith, D\BPBI M.%
\end{APACrefauthors}%
\unskip\
\newblock
\APACrefYearMonthDay{2024}{}{}.
\newblock
{\BBOQ}\APACrefatitle {Time Cells in the Retrosplenial Cortex} {Time cells in the retrosplenial cortex}.{\BBCQ}
\newblock
\APACjournalVolNumPages{bioRxiv}{}{}{}.
\PrintBackRefs{\CurrentBib}

\bibitem [\protect \citeauthoryear {%
Tank%
\ \BBA {} Hopfield%
}{%
Tank%
\ \BBA {} Hopfield%
}{%
{\protect \APACyear {1987}}%
}]{%
TankHopf87}
\APACinsertmetastar {%
TankHopf87}%
\begin{APACrefauthors}%
Tank, D.%
\BCBT {}\ \BBA {} Hopfield, J.%
\end{APACrefauthors}%
\unskip\
\newblock
\APACrefYearMonthDay{1987}{}{}.
\newblock
{\BBOQ}\APACrefatitle {Neural computation by concentrating information in time} {Neural computation by concentrating information in time}.{\BBCQ}
\newblock
\APACjournalVolNumPages{Proceedings of the National Academy of Sciences}{84}{7}{1896--1900}.
\PrintBackRefs{\CurrentBib}

\bibitem [\protect \citeauthoryear {%
Taxidis%
\ \protect \BOthers {.}}{%
Taxidis%
\ \protect \BOthers {.}}{%
{\protect \APACyear {2020}}%
}]{%
TaxiEtal20}
\APACinsertmetastar {%
TaxiEtal20}%
\begin{APACrefauthors}%
Taxidis, J.%
, Pnevmatikakis, E\BPBI A.%
, Dorian, C\BPBI C.%
, Mylavarapu, A\BPBI L.%
, Arora, J\BPBI S.%
, Samadian, K\BPBI D.%
\BDBL {}Golshani, P.%
\end{APACrefauthors}%
\unskip\
\newblock
\APACrefYearMonthDay{2020}{}{}.
\newblock
{\BBOQ}\APACrefatitle {Differential Emergence and Stability of Sensory and Temporal Representations in Context-Specific Hippocampal Sequences} {Differential emergence and stability of sensory and temporal representations in context-specific hippocampal sequences}.{\BBCQ}
\newblock
\APACjournalVolNumPages{Neuron}{108}{5}{984--998.e9}.
\PrintBackRefs{\CurrentBib}

\bibitem [\protect \citeauthoryear {%
Tenenbaum%
, de Silva%
\BCBL {}\ \BBA {} Langford%
}{%
Tenenbaum%
\ \protect \BOthers {.}}{%
{\protect \APACyear {2000}}%
}]{%
TeneEtal00}
\APACinsertmetastar {%
TeneEtal00}%
\begin{APACrefauthors}%
Tenenbaum, J\BPBI B.%
, de Silva, V.%
\BCBL {}\ \BBA {} Langford, J\BPBI C.%
\end{APACrefauthors}%
\unskip\
\newblock
\APACrefYearMonthDay{2000}{}{}.
\newblock
{\BBOQ}\APACrefatitle {A global geometric framework for nonlinear dimensionality reduction} {A global geometric framework for nonlinear dimensionality reduction}.{\BBCQ}
\newblock
\APACjournalVolNumPages{Science}{290}{5500}{2319-23}.
\newblock
\begin{APACrefDOI} \doi{10.1126/science.290.5500.2319} \end{APACrefDOI}
\PrintBackRefs{\CurrentBib}

\bibitem [\protect \citeauthoryear {%
Terada%
, Sakurai%
, Nakahara%
\BCBL {}\ \BBA {} Fujisawa%
}{%
Terada%
\ \protect \BOthers {.}}{%
{\protect \APACyear {2017}}%
}]{%
TeraEtal17}
\APACinsertmetastar {%
TeraEtal17}%
\begin{APACrefauthors}%
Terada, S.%
, Sakurai, Y.%
, Nakahara, H.%
\BCBL {}\ \BBA {} Fujisawa, S.%
\end{APACrefauthors}%
\unskip\
\newblock
\APACrefYearMonthDay{2017}{}{}.
\newblock
{\BBOQ}\APACrefatitle {Temporal and Rate Coding for Discrete Event Sequences in the Hippocampus} {Temporal and rate coding for discrete event sequences in the hippocampus}.{\BBCQ}
\newblock
\APACjournalVolNumPages{Neuron}{94}{}{1-15}.
\PrintBackRefs{\CurrentBib}

\bibitem [\protect \citeauthoryear {%
Tiganj%
, Cromer%
, Roy%
, Miller%
\BCBL {}\ \BBA {} Howard%
}{%
Tiganj%
\ \protect \BOthers {.}}{%
{\protect \APACyear {2018}}%
}]{%
TigaEtal18a}
\APACinsertmetastar {%
TigaEtal18a}%
\begin{APACrefauthors}%
Tiganj, Z.%
, Cromer, J\BPBI A.%
, Roy, J\BPBI E.%
, Miller, E\BPBI K.%
\BCBL {}\ \BBA {} Howard, M\BPBI W.%
\end{APACrefauthors}%
\unskip\
\newblock
\APACrefYearMonthDay{2018}{}{}.
\newblock
{\BBOQ}\APACrefatitle {Compressed timeline of recent experience in monkey {lPFC}} {Compressed timeline of recent experience in monkey {lPFC}}.{\BBCQ}
\newblock
\APACjournalVolNumPages{Journal of Cognitive Neuroscience}{30}{}{935-950}.
\PrintBackRefs{\CurrentBib}

\bibitem [\protect \citeauthoryear {%
Tiganj%
, Hasselmo%
\BCBL {}\ \BBA {} Howard%
}{%
Tiganj%
\ \protect \BOthers {.}}{%
{\protect \APACyear {2015}}%
}]{%
TigaEtal15}
\APACinsertmetastar {%
TigaEtal15}%
\begin{APACrefauthors}%
Tiganj, Z.%
, Hasselmo, M\BPBI E.%
\BCBL {}\ \BBA {} Howard, M\BPBI W.%
\end{APACrefauthors}%
\unskip\
\newblock
\APACrefYearMonthDay{2015}{}{}.
\newblock
{\BBOQ}\APACrefatitle {A Simple biophysically plausible model for long time constants in single neurons} {A simple biophysically plausible model for long time constants in single neurons}.{\BBCQ}
\newblock
\APACjournalVolNumPages{Hippocampus}{25}{1}{27-37}.
\PrintBackRefs{\CurrentBib}

\bibitem [\protect \citeauthoryear {%
Tsao%
\ \protect \BOthers {.}}{%
Tsao%
\ \protect \BOthers {.}}{%
{\protect \APACyear {2018}}%
}]{%
TsaoEtal18}
\APACinsertmetastar {%
TsaoEtal18}%
\begin{APACrefauthors}%
Tsao, A.%
, Sugar, J.%
, Lu, L.%
, Wang, C.%
, Knierim, J\BPBI J.%
, Moser, M\BHBI B.%
\BCBL {}\ \BBA {} Moser, E\BPBI I.%
\end{APACrefauthors}%
\unskip\
\newblock
\APACrefYearMonthDay{2018}{}{}.
\newblock
{\BBOQ}\APACrefatitle {Integrating time from experience in the lateral entorhinal cortex} {Integrating time from experience in the lateral entorhinal cortex}.{\BBCQ}
\newblock
\APACjournalVolNumPages{Nature}{561}{}{57-62}.
\PrintBackRefs{\CurrentBib}

\bibitem [\protect \citeauthoryear {%
Van~der Maaten%
\ \BBA {} Hinton%
}{%
Van~der Maaten%
\ \BBA {} Hinton%
}{%
{\protect \APACyear {2008}}%
}]{%
VanHint08}
\APACinsertmetastar {%
VanHint08}%
\begin{APACrefauthors}%
Van~der Maaten, L.%
\BCBT {}\ \BBA {} Hinton, G.%
\end{APACrefauthors}%
\unskip\
\newblock
\APACrefYearMonthDay{2008}{}{}.
\newblock
{\BBOQ}\APACrefatitle {Visualizing data using t-{SNE}} {Visualizing data using t-{SNE}}.{\BBCQ}
\newblock
\APACjournalVolNumPages{Journal of machine learning research}{9}{11}{}.
\PrintBackRefs{\CurrentBib}

\bibitem [\protect \citeauthoryear {%
Wallach%
, Melanson%
, Longtin%
\BCBL {}\ \BBA {} Maler%
}{%
Wallach%
\ \protect \BOthers {.}}{%
{\protect \APACyear {2022}}%
}]{%
WallEtal22}
\APACinsertmetastar {%
WallEtal22}%
\begin{APACrefauthors}%
Wallach, A.%
, Melanson, A.%
, Longtin, A.%
\BCBL {}\ \BBA {} Maler, L.%
\end{APACrefauthors}%
\unskip\
\newblock
\APACrefYearMonthDay{2022}{}{}.
\newblock
{\BBOQ}\APACrefatitle {Mixed selectivity coding of sensory and motor social signals in the thalamus of a weakly electric fish} {Mixed selectivity coding of sensory and motor social signals in the thalamus of a weakly electric fish}.{\BBCQ}
\newblock
\APACjournalVolNumPages{Current Biology}{32}{1}{51--63}.
\PrintBackRefs{\CurrentBib}

\bibitem [\protect \citeauthoryear {%
Ward%
, Tan%
\BCBL {}\ \BBA {} Grenfell-Essam%
}{%
Ward%
\ \protect \BOthers {.}}{%
{\protect \APACyear {2010}}%
}]{%
WardEtal10}
\APACinsertmetastar {%
WardEtal10}%
\begin{APACrefauthors}%
Ward, G.%
, Tan, L.%
\BCBL {}\ \BBA {} Grenfell-Essam, R.%
\end{APACrefauthors}%
\unskip\
\newblock
\APACrefYearMonthDay{2010}{}{}.
\newblock
{\BBOQ}\APACrefatitle {Examining the relationship between free recall and immediate serial recall: the effects of list length and output order} {Examining the relationship between free recall and immediate serial recall: the effects of list length and output order}.{\BBCQ}
\newblock
\APACjournalVolNumPages{Journal of Experimental Psychology: Learning, Memory and Cognition}{36}{5}{1207-41}.
\newblock
\begin{APACrefDOI} \doi{10.1037/a0020122} \end{APACrefDOI}
\PrintBackRefs{\CurrentBib}

\bibitem [\protect \citeauthoryear {%
Wei%
\ \BBA {} Stocker%
}{%
Wei%
\ \BBA {} Stocker%
}{%
{\protect \APACyear {2012}}%
}]{%
WeiStoc12}
\APACinsertmetastar {%
WeiStoc12}%
\begin{APACrefauthors}%
Wei, X\BHBI X.%
\BCBT {}\ \BBA {} Stocker, A\BPBI A.%
\end{APACrefauthors}%
\unskip\
\newblock
\APACrefYearMonthDay{2012}{}{}.
\newblock
{\BBOQ}\APACrefatitle {Bayesian inference with efficient neural population codes} {Bayesian inference with efficient neural population codes}.{\BBCQ}
\newblock
\BIn{} \APACrefbtitle {Artificial Neural Networks and Machine Learning--ICANN 2012} {Artificial neural networks and machine learning--icann 2012}\ (\BPGS\ 523--530).
\newblock
\APACaddressPublisher{}{Springer}.
\PrintBackRefs{\CurrentBib}

\bibitem [\protect \citeauthoryear {%
Weyl%
}{%
Weyl%
}{%
{\protect \APACyear {1922}}%
}]{%
Weyl22}
\APACinsertmetastar {%
Weyl22}%
\begin{APACrefauthors}%
Weyl, H.%
\end{APACrefauthors}%
\unskip\
\newblock
\APACrefYear{1922}.
\newblock
\APACrefbtitle {Space--time--matter} {Space--time--matter}.
\newblock
\APACaddressPublisher{}{Dutton}.
\PrintBackRefs{\CurrentBib}

\bibitem [\protect \citeauthoryear {%
K.~Zhang%
}{%
K.~Zhang%
}{%
{\protect \APACyear {1996}}%
}]{%
Zhan96}
\APACinsertmetastar {%
Zhan96}%
\begin{APACrefauthors}%
Zhang, K.%
\end{APACrefauthors}%
\unskip\
\newblock
\APACrefYearMonthDay{1996}{}{}.
\newblock
{\BBOQ}\APACrefatitle {Representation of spatial orientation by the intrinsic dynamics of the head-direction cell ensemble: a theory.} {Representation of spatial orientation by the intrinsic dynamics of the head-direction cell ensemble: a theory.}{\BBCQ}
\newblock
\APACjournalVolNumPages{Journal of Neuroscience}{16}{6}{2112-26}.
\PrintBackRefs{\CurrentBib}

\bibitem [\protect \citeauthoryear {%
W.~Zhang%
, Wu%
\BCBL {}\ \BBA {} Wu%
}{%
W.~Zhang%
\ \protect \BOthers {.}}{%
{\protect \APACyear {2022}}%
}]{%
ZhanEtal22b}
\APACinsertmetastar {%
ZhanEtal22b}%
\begin{APACrefauthors}%
Zhang, W.%
, Wu, Y\BPBI N.%
\BCBL {}\ \BBA {} Wu, S.%
\end{APACrefauthors}%
\unskip\
\newblock
\APACrefYearMonthDay{2022}{}{}.
\newblock
{\BBOQ}\APACrefatitle {Translation-equivariant representation in recurrent networks with a continuous manifold of attractors} {Translation-equivariant representation in recurrent networks with a continuous manifold of attractors}.{\BBCQ}
\newblock
\APACjournalVolNumPages{Advances in Neural Information Processing Systems}{35}{}{15770--15783}.
\PrintBackRefs{\CurrentBib}

\bibitem [\protect \citeauthoryear {%
Zuo%
\ \protect \BOthers {.}}{%
Zuo%
\ \protect \BOthers {.}}{%
{\protect \APACyear {2024}}%
}]{%
ZuoEtal23}
\APACinsertmetastar {%
ZuoEtal23}%
\begin{APACrefauthors}%
Zuo, S.%
, Wang, C.%
, Wang, L.%
, Jin, Z.%
, Kusunoki, M.%
\BCBL {}\ \BBA {} Kwok, S\BPBI C.%
\end{APACrefauthors}%
\unskip\
\newblock
\APACrefYearMonthDay{2024}{}{}.
\newblock
{\BBOQ}\APACrefatitle {Neural signatures for temporal-order memory in the medial posterior parietal cortex} {Neural signatures for temporal-order memory in the medial posterior parietal cortex}.{\BBCQ}
\newblock
\APACjournalVolNumPages{bioRxiv}{}{}{2023--08}.
\PrintBackRefs{\CurrentBib}

\end{thebibliography}

\newpage
\subsection*{Supporting Information Appendix (SI)}
\setcounter{figure}{0}
\renewcommand{\figurename}{Fig.}
\renewcommand{\thefigure}{S\arabic{figure}}

\begin{figure*}[hb]
\begin{center}
\includegraphics[width=0.99\textwidth]{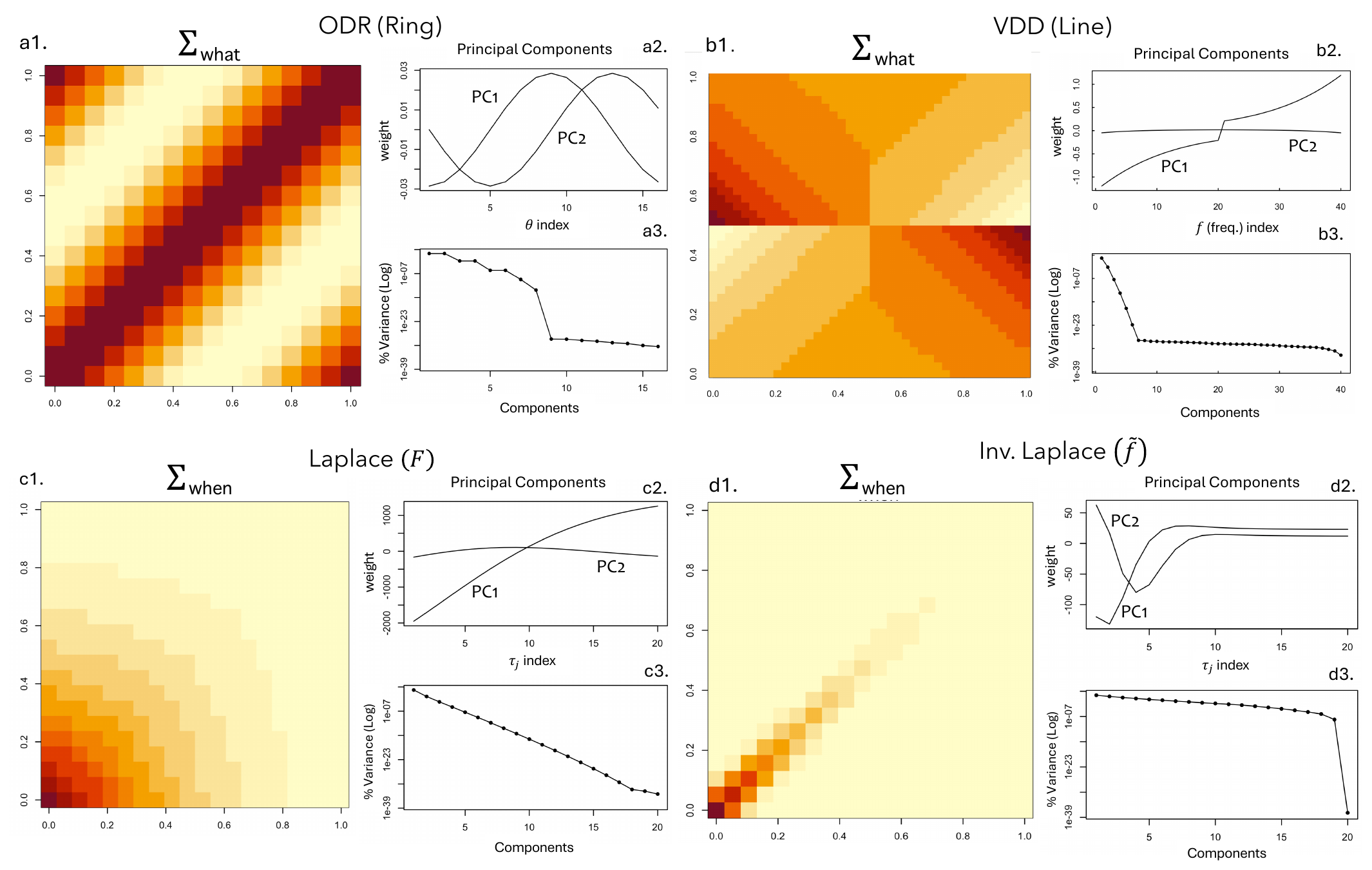}
\end{center}
\caption{ \emph{Covariance Matrices and Principal Components for \covwhat (top) and \covwhen (bottom), with the variance explained by the components shown in a log scale. }
    \textit{Top:} The stimulus covariance matrix (\covwhat) shown for the ODR (\textit{a1}, \textit{Left}) and VDD (\textit{a2}, \textit{Right}) tasks, where the task variable lie on a ring and a line, respectively. The circular task variable is tiled with circular Gaussian (Von Mises) receptive fields with periodic boundary conditions, producing a periodic covariance matrix, which generates two sinusoidal principal components (\textit{a2}) explaining the most variance (\textit{a3}). The line variable is tiled with sets of exponentially decaying and ramping cells, which generates distinct quadrants in the covariance matrix (\textit{b1}), with a \textit{smoothly-varying} first principal component (\textit{b2}) (with a discontinuity when switching from decaying to ramping cells) which explains the most variance (\textit{b3}), and a mostly flat second component.     
	 \textit{Bottom:} The time covariance matrix (\covwhen) shown for Laplace (temporal context cells, $F$) and Inverse Laplace (time cells, $\tilde{f}$) populations. With exponentially decaying cells, the covariance matrix for Laplace neurons (\textit{c1}) has a smoothly decaying activation across the diagonal, which is the direction of maximum variance. The first principal component picks this up to show a \textit{monotonically-varying} trend (\textit{c2}) and also explains the most variance (\textit{c3}). Time cells on the other hand are sequentially firing cells tiling a time-line, which shows up as a diagonally-concentrated covariance matrix which decays from bottom right to top left (\textit{d1}). The biggest principal components (\textit{d2}) of this covariance matrix also end up being oscillating sinusoids (similar to \textit{a2}) but with a decaying envelope as we go from left to right (smaller to larger $\tau_j$), and explain the most variance (\textit{d3}). For \textit{both} these temporal representations, the variance explained (set on a \textit{logarithmic} scale) with number of principal components goes down smoothly, underscoring the \textit{high-dimensional} nature of the covariance matrices (\textit{c3} and \textit{d3}).}\label{fig:supp}
\end{figure*}

\end{document}